\documentclass[fleqn,usenatbib]{mnras}

\usepackage[T1]{fontenc}

\DeclareRobustCommand{\VAN}[3]{#2}
\let\VANthebibliography\thebibliography
\def\thebibliography{\DeclareRobustCommand{\VAN}[3]{##3}\VANthebibliography}

\usepackage{graphicx}	
\usepackage{amsmath}	
\usepackage{amssymb}	
\usepackage{eso-pic}
\usepackage{pdflscape}
\usepackage{longtable}
\usepackage{threeparttablex}
\usepackage{hyperref}
\usepackage{bm}

\usepackage{newtxtext,newtxmath}

\title[Scalable hierarchical BayeSN inference]{Scalable hierarchical BayeSN inference: Investigating dependence of SN Ia host galaxy dust properties on stellar mass and redshift}
\author[M. Grayling et al.]{
Matthew Grayling$^{1}$\thanks{Contact e-mail: \href{mailto:mg2102@cam.ac.uk}{mg2102@cam.ac.uk}},
Stephen Thorp$^{2}$,
Kaisey~S.~Mandel$^{1, 3}$, 
Suhail Dhawan$^{1}$, 
Ana Sofia M. Uzsoy$^{4}$, 
\newauthor
\ Benjamin M. Boyd$^{1}$, 
Erin E. Hayes$^{1}$ 
and Sam M. Ward$^{1}$
\\
$^{1}$Institute of Astronomy and Kavli Institute for Cosmology, Madingley Road, Cambridge CB3 0HA, UK \\
$^{2}$Oskar Klein Centre, Department of Physics, Stockholm University, SE-106 91 Stockholm, Sweden \\
$^{3}$Statistical Laboratory, DPMMS, University of Cambridge, Wilberforce Road, Cambridge, CB3 0WB, UK \\
$^{4}$Center for Astrophysics | Harvard \& Smithsonian, Cambridge, MA 02138, USA
}

\date{Accepted XXX. Received YYY; in original form ZZZ}

\pubyear{2023}

\begin{document}
\label{firstpage}
\pagerange{\pageref{firstpage}--\pageref{lastpage}}
\maketitle

\begin{abstract}
We apply the hierarchical probabilistic SED model BayeSN to analyse a sample of 475 SNe Ia (0.015 < z < 0.4) from Foundation, DES3YR and PS1MD to investigate the properties of dust in their host galaxies. We jointly infer the dust law $R_V$ population distributions at the SED level in high- and low-mass galaxies simultaneously with dust-independent, intrinsic differences. We find an intrinsic mass step of $-0.049\pm0.016$ mag, at a significance of 3.1$\sigma$, when allowing for a constant intrinsic, achromatic magnitude offset. We additionally apply a model allowing for time- and wavelength-dependent intrinsic differences between SNe Ia in different mass bins, finding $\sim$2$\sigma$ differences in magnitude and colour around peak and 4.5$\sigma$ differences at later times. These intrinsic differences are inferred simultaneously with a difference in population mean $R_V$ of $\sim$2$\sigma$ significance, demonstrating that both intrinsic and extrinsic differences may play a role in causing the host galaxy mass step. We also consider a model which allows the mean of the $R_V$ distribution to linearly evolve with redshift but find no evidence for any evolution - we infer the gradient of this relation $\eta_R = -0.38\pm0.70$. In addition, we discuss in brief a new, GPU-accelerated Python implementation of BayeSN suitable for application to large surveys which is publicly available and can be used for future cosmological analyses; this code can be found here: \url{https://github.com/bayesn/bayesn}.
\end{abstract}




\section{Introduction}
\label{intro}


Type Ia supernovae (SNe~Ia) are bright, thermonuclear explosions of carbon-oxygen white dwarfs in a binary system \citep[e.g., see][]{Maguire2017}. They are excellent distance indicators, were used for the discovery of the accelerated expansion of the universe \citep{Riess:1998cb,Perlmutter:1998np}, and play a central role in precisely constraining the properties of dark energy, for example its equation of state ($w$) and time dependence ($w_a$) \citep[][]{Brout22, DES24}. SNe~Ia are also crucial for local measurements of the Hubble Constant ($H_0$) via the distance ladder \citep{Riess2022}, which is currently in $\sim 5\sigma$ tension with the inference from the early universe \citep{Planck20}. 

While the current best constraints on $w$ are dominated by statistical errors \citep{Vincenzi24}, the order-of-magnitude larger samples expected from upcoming surveys such as the Legacy Survey of Space and Time at Vera Rubin Observatory \citep[LSST; ][]{Ivezic19} -- as well as complementary low-redshifts surveys such as the Young Supernova Experiment \citep[YSE; ][]{Jones21, Aleo23} and Zwicky Transient Facility \citep[ZTF; ][]{Bellm19b, Bellm19a} -- means obtaining more precise cosmological constraints will increasingly rely on better understanding of systematics. Improving the constraints on cosmological parameters from SNe Ia hinges on improved physical understanding of the nature of these events. In this paper, we present analysis of a combined sample of 475 SNe Ia using the hierarchical Bayesian spectral energy distribution (SED) model BayeSN \citep{M20, T21}, focusing on how their properties vary as a function of both host galaxy stellar mass and redshift. In addition, we present a new GPU-accelerated Python implementation of BayeSN which we make publicly available.

A key open question within the field at present is the nature of the host galaxy `mass step', whereby supernovae in higher mass galaxies are on average brighter than those in lower mass galaxies after standardisation \citep{Sullivan10, Kelly10}. This effect has been consistently observed in optical samples with a mass step of ${\sim}0.04-0.1$ mag \citep[e.g.][]{Kelly10, Sullivan10, Lampeitl10, Gupta11, Childress13, Betoule14, Jones18, Jones19, Roman18, Smith20, Kelsey21}, with the split between high- and low-mass hosts typically placed between $10^{10}$ M$_\odot$ \citep[e.g.][]{Sullivan10} and $10^{10.8}$ M$_\odot$ \citep[e.g.][]{Kelly10}. Relations have also been demonstrated between SN luminosity and other environmental properties, including star formation rate (SFR), specific star formation rate (sSFR) and rest-frame colour \citep[e.g.][]{Lampeitl10, Sullivan10, Briday22}. Subsequent studies have demonstrated that SN luminosity often show a stronger relation with local properties, derived from the region in which each SN exploded rather than from the whole galaxy \citep[e.g.][]{Rigault13, Rigault15, Rigault20, Jones15, Moreno-Raya16b, Moreno-Raya16a, Jones18, Kim18, Roman18, Kim19, Rose19, Kelsey21}. For example, \citet{Rigault20} found a step of 0.163$\pm$0.029 mag when looking at local sSFR, compared with a mass step of 0.119$\pm$0.032 mag on the same sample. \citet{Kelsey21} found that the maximum step size increased by $\sim40$ per cent when considering the mass enclosed in an aperture around the SN position rather than the total mass of the host galaxy.

The exact physical cause of the mass step remains open for debate. It could, for example, be a result of intrinsic differences in the SN populations in high- and low-mass galaxies; a number of works have posited that this effect could result from a difference in SN progenitors in different environments \citep[e.g.][]{Sullivan10, Childress14, Kim18, Rigault20, Briday22}. Alternatively, the mass step could arise from extrinsic differences; \citet{BS21} proposed that it can be explained entirely by differences in host galaxy dust properties between high- and low-mass SN hosts, requiring no difference in intrinsic SN populations.

Several works have found results consistent with the view that extrinsic effects can explain the mass step. The aforementioned study presented in \citet{BS21} finds that a dust parameter $R_V$ distribution with a mean $\mu_R$ of $1.50\pm0.25$ for high-mass hosts and $2.75\pm0.35$ for low-mass hosts, along with a wide standard deviation of $\sigma_R=1.3\pm0.2$, provides the best match to the \textsc{SALT2} \citep{Guy07} fits of 1445 SNe Ia  from the Carnegie Supernova Project I \citep[CSP-I; ][]{Contreras10, Strizinger11, Krisciunas17}, the CfA \citep{Hickens09, Hickens12}, Foundation \citep{Foley18, Jones19}, the Pan-STARRS-1 Medium Deep Survey \citep{Rest14, Scolnic18}, Supernova Legacy Survey \citep{Astier06, Betoule14}, SDSS-II \citep{Frieman08, Sako11, Sako18} and the Dark Energy Survey \citep[DES; ][]{DES, Brout19}. \citet{Popovic23} updates the methods of \citet{BS21} to perform inference using an MCMC sampler, finding $\mu_R$ for high- and low-mass hosts of $2.138\pm0.25$ and $3.026\pm0.375$ respectively for the Pantheon+ sample of 1701 SNe Ia \citep{Brout22}. \citet{Wiseman23} finds that a model with a galaxy age-varying $R_V$ and no intrinsic luminosity difference replicates the observed SN population well, although does not rule out that intrinsic differences may play a role in the mass step. \citet{Meldorf23} analysed the dust laws of 1100 SN host galaxies in DES and found a significant difference in $R_V$ between high- and low-mass galaxies of $\sim$0.7, albeit this was based on analysis of the attenuation law of the full galaxy rather than focusing on extinction along the line-of-sight to SN positions.

In contrast, a number of other works have instead been consistent with the view that the mass step results at least in part from intrinsic differences. For example, \citet{Gonzalez-Gaitan21} finds evidence for a mass step alongside a difference in $R_V$ values between high- and low-mass galaxies, and posits that this may result from a difference in intrinsic colour. In a separate analysis of attenuation laws of SN host galaxies in DES, \citet{Duarte23} found that host galaxy dust differences cannot fully explain the mass step. \citet{Jones23} presented a version of the SALT3 model \citep{Kenworthy21} which incorporates a relation between host galaxy stellar mass and the SN Ia SED, finding evidence for differences in spectroscopic and photometric properties of SNe Ia between each bin, although in the latter case the model could not discern intrinsic differences from those caused by dust. \citet{Taylor24} applied the SALT3 model to two separate samples split based on host galaxy stellar mass, also finding some differences between the two although not discerning between intrinsic colour and dust reddening. Recent analysis of the DES-SN 5YR sample of SNe Ia from DES \citep{DES24} found inconsistencies between parameters inferred from simulations based on a dust-only mass step and those inferred from real data, suggesting that dust alone cannot explain the mass step \citep{Vincenzi24}.

Additionally, if the mass step were solely a result of host galaxy dust, one would expect that no mass step would be observed in near-infrared (NIR) wavelengths where dust -- particularly the value of $R_V$ -- has little effect. Despite this, several works have found evidence for a mass step in NIR wavelengths. \citet{Ponder21} analysed 143 SNe Ia in $H$-band and found a significant step of $0.13\pm0.04$ at a mass of $10^{10.43}M_\odot$; omitting notable outliers, the most significant step is $0.08\pm0.04$ at $10^{10.65}M_\odot$. \citet{Uddin20} analysed a sample of 113 SNe Ia from CSP-I, fitting them using the \textsc{max\_model} method within \textsc{SNoopy} \citep{Burns11} across $uBgVriYJH$ bands. Based on a split at the median stellar mass of the sample, $10^{10.65}M_\odot$, they find an $H$-band step of $0.093\pm0.043$ mag and a $J$-band step of $0.090\pm0.043$ mag - overall, they find NIR steps of comparable significance to the optical steps and that the relation between step size and wavelength was inconsistent with a dust-driven step. \citet{Jones22} analysed a sample of 79 SNe, including 42 low redshift SNe (z < 0.1) from CSP-I and 37 higher redshift (0.2 < z < 0.7) SNe with rest-frame NIR, and found a mass step from NIR light curves of $0.072\pm0.041$ mag at mass of $10^{10.44}M_\odot$.

However, it should be noted that other NIR analyses have not found evidence for a mass step. \citet{Johansson21} analysed a sample of $\sim$240 SNe Ia using the \textsc{colour\_model} method within \textsc{SNooPy}, assuming a fixed host galaxy $R_V=2.0$. This study finds NIR mass steps in $J$ and $H-$bands of $0.021\pm0.033$ mag and $-0.020\pm0.036$ respectively, both consistent with zero, although not inconsistent with the steps found in \citet{Uddin20} considering the uncertainties. They also find the mass step to be consistent with zero across optical and NIR bands when allowing $R_V$ to be a free parameter for each SN.

Most recently, \citet{Uddin23} analysed a sample of 325 SNe from CSP-I and CSP-II \citep{Phillips19}, including the sample of 113 SNe analysed in \citet{Uddin20}. Using a more recent version of the \textsc{max\_model} method within \textsc{SNooPy}, they find no significant mass step in any optical or NIR band, with the possible exception of $u$-band which shows a step of $-0.151\pm0.069$ mag. They also demonstrate that their inferred correlations between host mass and luminosity do not vary with wavelength. Both of these findings are consistent whether using only the CSP-I sample from \citet{Uddin20} or the combined CSP-I and CSP-II sample. Additionally, \citet{Karchev23b} applied Bayesian model comparison using simulation-based inference (SBI) and found results disfavouring both intrinsic and extrinsic differences.

The debate between a mass step driven by extrinsic effects and intrinsic variation in the SN population closely relates to a general challenge in SNe Ia -- disentangling the intrinsic distribution from the effects of host galaxy dust extinction. It is challenging, for example, to tell apart an intrinsically red SN from an intrinsically typical SN which happened to explode in a dusty environment. \textsc{SALT} \citep{Guy07, Kenworthy21} is the most widely used model for SN Ia analyses, however a key limitation is that it compresses all colour information into a single $c$ parameter, which roughly corresponds to peak $B-V$ apparent colour. As demonstrated in \citet{Mandel17}, with this approach care must be taken to avoid confounding intrinsic variation with dust. While some \textsc{SALT}-based analyses do break down the $c$ parameter into separate intrinsic and dust components \citep[e.g.][]{Mandel17, BS21, Popovic23}, \textsc{SALT} is trained with a single colour law describing the effect that the apparent colour parameter $c$ has on the underlying SN SED, meaning that it cannot accommodate the possibility of different dust laws when fitting observed light curves. 

Unlike \textsc{SALT}, \textsc{SNooPy} does include an explicit treatment of dust extinction. However, in addition to host galaxy dust it is also important to incorporate the intrinsic colour distribution of SNe, including variations beyond those correlated with light curve shape. Excluding this effect can bias estimates of $R_V$ \citep{Mandel17} and lead to overestimates of the colour variation caused by an $R_V$ distribution \citep{TM22}.

The challenge of disentangling dust from intrinsic SN variation has led to the development of hierarchical Bayesian models for analysing samples of SNe Ia. Hierarchical modelling was first applied to this end in \citet{Mandel09, Mandel11} and again in \citet{Burns14}; the former method has since been developed further into the hierarchical SED model BayeSN \citep{T21,M20}. The advantage of a hierarchical approach is that it allows for joint inference of population level and individual SN parameters. Specific to this problem, it allows for explicit, separate treatments of the intrinsic and host galaxy dust distributions of SNe Ia, with the population-level parameters inferred while marginalising over all individual SN (latent) parameters. Hierarchical modelling also provides more precise inference of the width of a population, when compared with simply taking the standard deviation of a number of individual estimates; the latter approach will lead to overestimates as it fails to take into account errors on the individual estimates \citep[see discussion in Section 5 of][]{TM22}. \citet{Wojtak23} presents a recent analysis of SNe Ia using hierarchical Bayesian modelling, finding evidence for two separate populations although not splitting the population directly based on host galaxy stellar mass.

Both \citet{T21} and \citet{TM22} apply BayeSN to study the dust distributions of SNe in high- and low-mass hosts and find no evidence for a difference in $R_V$ distributions between different environments. Notably, \citet{TM22} finds evidence for a NIR mass step even when allowing for different dust distributions between high- and low-mass hosts.

In this work, we use a new GPU-accelerated implementation of BayeSN which has increased performance by factors of $\sim$50-100, making large-scale hierarchical analyses of SNe Ia more than feasible; we make this code publicly available. We build on previous analysis in \citet{T21} in applying BayeSN to optical data to study the host galaxy dust properties of SNe Ia. We apply the model trained in \citet{T21} to higher redshift SNe from the Dark Energy Survey three-year sample (DES3YR) and the Pan-STARRS Medium Deep survey (PS1MD), increasing both the sample size and redshift range covered. We investigate the host galaxy dust distributions in both high- and low-mass hosts in tandem with dust-independent intrinsic differences, allowing us to comment on the debate between intrinsic and extrinsic causes. We also perform the first hierarchical Bayesian analysis with a redshift-dependent $R_V$ distribution to test whether there is any evidence that SN host galaxy dust properties evolve with redshift.

In Section \ref{data}, we detail the SN samples used in this work and describe redshift cuts applied to mitigate for selection effects. In Section \ref{method}, we summarise the BayeSN model and our new GPU-accelerated implementation, as well as detailing the specific models and analysis variants used within this work. We then describe our results, looking separately at a single dust population (Section \ref{Rv_pop_results}), multiple dust populations split by host galaxy stellar mass (Section \ref{Rv_split_results}) and a redshift-dependent dust distribution (Section \ref{Rv_z_results}), before concluding in Section \ref{conclusion}.

\section{Data}
\label{data}

For this work we analyse three separate spectroscopically-confirmed optical samples of SNe Ia observed in $griz$ bands: the Foundation DR1 sample \citep{Foley18}, the Dark Energy Survey \citep[DES;][]{DES} 3-year sample \citep[DES3YR;][]{Brout19, Smith20} and the sample from the Pan-STARRS Medium Deep survey \citep[PS1MD;][]{Kaiser10} as compiled in the Pantheon sample \citep{Scolnic18}. For all three samples, we use photometric data taken from the Pantheon+ compilation \citep{PantheonPlus}, which corrects for cross-calibration systematics between surveys. We include SNe from PS1MD present in the Pantheon compilation, rather than the Pantheon+ compilation, to increase our sample size as there were a few objects which were removed in the latter compilation that are suitable for our analysis.

All of the SNe in this analysis pass standard cosmology cuts \citep[e.g.][]{Scolnic18, PantheonPlus} and have global host galaxy masses measured from SED fits to $ugrizy$ and $ugriz$ photometry for Foundation \citep{Jones19} and PS1MD \citep{Scolnic18}, and from DES $griz$ photometry for DES3YR \citep{Brout19, Smith20, Wiseman20}.

\begin{figure}
\centering
\includegraphics[width = \linewidth]{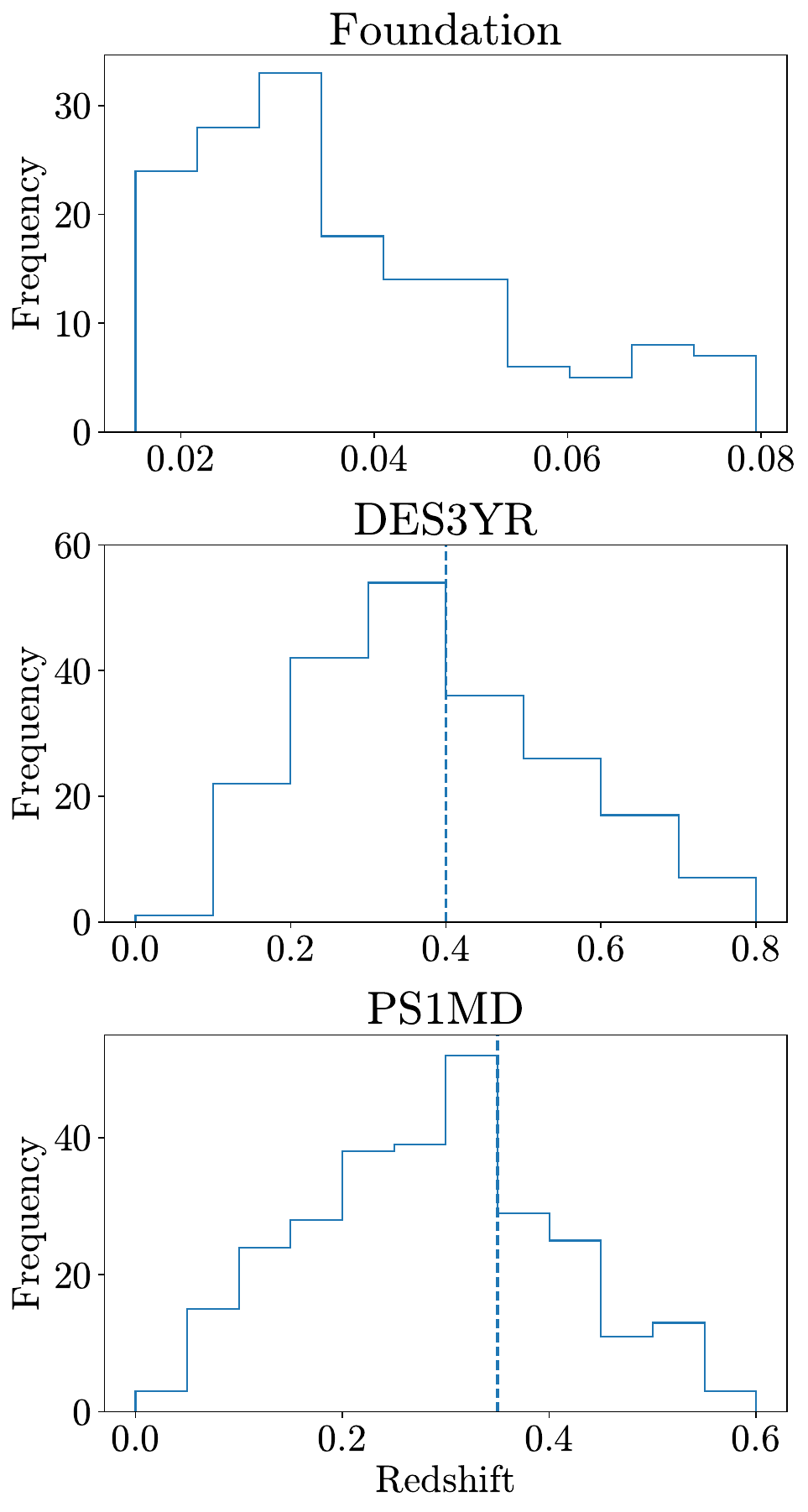}
\caption{Redshift distributions for the 3 samples used in this work; Foundation, DES3YR and PS1MD. Dashed lines for DES3YR and PS1MD indicate the upper redshift limit of the volume limited sub-samples included in our analysis.}
\label{z_dists}
\end{figure}

Within our analysis, we take care to mitigate for selection effects such as Malmquist bias \citep{Malmquist22}. Foundation DR1 is a local sample; we apply the same selection cuts as used in \citet{T21} for a total of 157 SNe Ia which span a redshift range of 0.015 < z < 0.08, where the sample has minimal impact from Malmquist bias \citep[][see further discussion in Appendix \ref{selection_effects}]{Foley18}. In contrast, DES3YR and PS1MD are higher redshift meaning that they are more susceptible to impact from selection effects. DES3YR comprises 207 SNe Ia over a redshift range of 0.077 < z < 0.85, while PS1MD comprises 279 SNe Ia over a range of 0.026 < z < 0.63. To mitigate for this, in this work we take volume limited subsets of the full samples - we apply a cut at the point where the redshift distribution begins to decrease, indicating that a large number of SNe have been missed due to selection effects. For DES3YR, we apply a redshift cut at 0.4 to give a sample of 119 SNe Ia, while for PS1MD we apply a cut at 0.35 for a final sample of 199 SNe Ia. When combined, these samples include a total of 475 SNe Ia. Figure \ref{z_dists} shows the redshift distributions of the three samples used in this work, with dashed lines indicating the upper redshift cuts applied to DES3YR and PS1MD. For discussion of these redshift cuts, please see Appendix \ref{selection_effects}. In brief, we investigate how the total $V$-band extinction varies with redshift and find no evidence that these volume limited sub-samples are significantly impacted by selection effects, although cannot rule out that some small effects may remain.

It should be noted that in this work we utilise the version of the BayeSN model presented in \citet{T21}, which is defined down to a rest-frame wavelength of $3500$ \AA, while the redshift range spanned by DES3YR and PS1MD means that rest-frame $g$-band often covers wavelengths below this lower limit. As such, for the majority of SNe in these two samples we disregard $g$-band data.

\section{Method}
\label{method}

For this work, we utilise the hierarchical Bayesian SN Ia SED model BayeSN  \citep{T21,M20}. Hierarchical Bayes provides an ideal framework for joint inference of populations and the individual objects which comprise them, for example looking in tandem at individual SN host galaxies and the overall population. In this section we detail the model used for our analysis and discuss the different variants used.

\subsection{Numpyro Implementation of BayeSN}
\label{numpyro}

Compared with previous BayeSN analyses \citep[e.g.][]{M20, T21, TM22, W22, Dhawan23} which implemented the model in \textsc{Stan} \citep{Carpenter17, stan}, we use a version implemented using the probabilistic programming Python library \textsc{numpyro} \citep{Phan19, Bingham19}. All computations in \textsc{numpyro} are carried out using \textsc{jax} \citep{jax}, meaning that this implementation of BayeSN supports just-in-time (JIT) compilation and GPUs, and similarly to the \textsc{Stan} version is fully differentiable with autodiff. By combining vectorised likelihood evaluation across the SN population with the use of GPU acceleration, the speed of the inference has increased by factors of $\sim$50-100, enabling hierarchical analyses of much larger SN samples. The \textsc{numpyro} implementation of the BayeSN model can be found here: \url{https://github.com/bayesn/bayesn}.

BayeSN uses the \textsc{numpyro} implementation of the No U-Turn Sampler \citep[NUTS;][]{Hoffman11}, an adaptive variant of Hamiltonian Monte Carlo \citep[HMC;][]{Neal11, Betancourt17}, to sample from target posterior distributions. To give an example of typical performance, for the hierarchical SED- and population-level dust models presented in this work HMC typically converges and generates sufficient effective sample sizes in $\sim$15-30 minutes; these values are quoted for 4 chains run in parallel across 4 NVIDIA A100 GPUs, using resources from the Cambridge Service for Data Driven Discovery (CSD3). This runtime allows for quick exploration of different model variants. 

While this work focuses on the physical properties of the SN Ia population and does not touch on cosmological analysis (as discussed in Section \ref{distcond} our results are conditioned on a fixed cosmology), the BayeSN code we make available can also be used to infer distance. A model is first `trained' by inferring the population-level parameters of the model \citep[as in][]{M20, T21, W22}. These population-level parameters are then fixed and the model can be applied to infer a cosmology-independent distance modulus $\mu^s$ simultaneously with all the other latent SN parameters, conditioned on those fixed population parameters. Using this approach, distance is jointly inferred with SN latent parameters rather than being estimated via the Tripp formula \citep{Tripp98} -- a post-hoc linear relation with those SN parameters -- as in SALT-based analyses. The advantage of this method is that it allows the full SN light curve to be used when estimating distance rather than compressing it to a single colour at peak, and also provides a distance estimate separately marginalised over the intrinsic and extrinsic variation observed in the population. Work is underway to integrate our code within the SN analysis framework SNANA \citep{Kessler09}, and future cosmological analyses will be able to use BayeSN-derived distances.

\subsection{The BayeSN Model}
\label{bayesn_model}

BayeSN is an SED model for SNe Ia, with the full time- and wavelength-varying SED given by,
\begin{equation}
\label{bayesn_equation}
\begin{aligned}
    -2.5\log_{10}[S_s(t,\lambda_r)/S_0(t,\lambda_r)] = M_0 + W_0(t,\lambda_r) \ + \ \\ \delta M^s+\theta^s_1W_1(t,\lambda_r) + \epsilon^s(t,\lambda_r)+A^s_V\xi\big(\lambda_r;R^{(s)}_V\big)
\end{aligned}
\end{equation}
where $t$ is the rest-frame phase relative to B-band maximum and $\lambda_r$ is rest-frame wavelength. $S_0(t, \lambda_r)$ is the optical-NIR SN Ia SED template of \citet{Hsiao07} and is fixed a priori as a zeroth-order template, along with the normalisation constant $M_0$ which is set to $-19.5$. All parameters denoted with $s$ are latent SN parameters, with unique values for each SN $s$, while all other parameters are global hyperparameters shared across the population. The different components which make up the model are described as follows:

\begin{itemize}
    \item $W_0(t, \lambda_r)$ is a function that warps and normalises the zeroth-order SED template to create a mean intrinsic SED for the population, while $W_1(t, \lambda_r)$ is a functional principal component (FPC) that describes the first mode of intrinsic SED variation for SNe Ia. Both of these are implemented as cubic spline surfaces.
    \item $\theta_1^s$ is a coefficient quantifying the effect of the $W_1$ FPC for each SN. This is modelled as a normal distribution, with $\theta_1^s \sim N(0, 1)$\footnote{Throughout this paper, we use the notation $x\sim N(\mu, \sigma^2)$ where $\mu$ is the mean and $\sigma^2$ is the variance.}. Combined, $W_1(t, \lambda_r)$ and $\theta_1^s$ capture the `broader-brighter' relation observed in SNe Ia where intrinsically brighter light curves evolve over longer timescales around peak \citep{Phillips93}. The parameter $\theta_1^s$ can therefore be thought of as analogous to a stretch parameter as present in models such as SALT and SNooPy.
    \item $\delta M^s$ is an achromatic, time-independent magnitude offset for each SN, drawn from a normal distribution with $\delta M^s \sim N(0, \sigma_0^2)$. $\sigma_0$ is a hyperparameter describing the width of this distribution and is inferred when the model is trained.
    \item $\epsilon^s(t,\lambda_r)$ is a time- and wavelength-dependent function that describes the time-varying residual intrinsic colour variations in the SED not captured by $\theta_1^sW_1(t, \lambda_r)$. This parameter is represented by a cubic spline function over time and wavelength defined by a matrix of knots $\mathbf{E}^s$. These are drawn from a multivariate Gaussian $\mathbf{e}^s \sim N(0, \mathbf{\Sigma}_\epsilon$), where $\mathbf{e}^s$ is the vectorisation of the $\mathbf{E}^s$ matrix. The covariance matrix $\mathbf{\Sigma}_\epsilon$ is a hyperparameter of the model which describes the distribution of residual scatter across the population of SNe Ia and is inferred during training.
    \item $A_V^s$ and $R_V^{(s)}$ describe the host galaxy extinction law for each SN; $A_V^s$ is the total amount of V-band extinction, while $R_V^{(s)}$ parameterises the \citet{Fitzpatrick99} dust extinction law assumed by the model, here denoted by $\xi\big(\lambda_r;R^{(s)}_V\big)$. $R_V^{(s)}$ can be treated as either a shared parameter across the population or a unique value for each SN (see Section \ref{Rv_distributions} for discussion of different assumed $R_V^s$ distributions). For $A_V^s$, we assume an exponential prior described by a scale parameter (the population mean) $\tau_A$; that is, $A_V \sim \text{Exponential}(\tau_A)$ with probability density $P(A_V | \tau_A) = \tau_A^{-1}\, \exp(-A_V / \tau_A)$ for $A_V \ge 0$, and zero otherwise.
\end{itemize}

The rest-frame, host galaxy dust-extinguished SED model $S_s(t, \lambda_r)$ is then scaled based on distance modulus $\mu^s$, redshifted and corrected for Milky Way dust extinction using a \citet{Fitzpatrick99} dust law with $R_V=3.1$ and dust maps from \citet{Schlafly11}. This produces an observer-frame SED, which can be integrated through photometric filters to produce model photometry that can in turn be compared with observed photometry to compute a likelihood. This hierarchical model is trained by inferring all global parameters and latent parameters for a population of SNe Ia 
(see \citet{M20} and \citet{T21} for full discussion of model training). The latent parameters are marginalised over and posterior estimates are obtained for the global parameters.

Previous SALT-based analyses which aimed to disentangle the intrinsic colour of SNe Ia from dust reddening \citep[e.g.][]{Mandel17, BS21, Popovic21} have separated out the typical linear relationship between peak $B-V$ colour and peak magnitude, characterised by a slope $\beta$, into an intrinsic colour relation with slope $\beta_\text{int}$ and dust relation with slope $R_B$. Within the BayeSN model, there is a degree of colour dependence on $\theta_1$ i.e. the stretch-like parameter. The dependence of the SED on the component of colour which is independent of both dust and stretch is parameterised by $\epsilon^s$, which as discussed above is drawn from a multivariate Gaussian $\mathbf{e}^s \sim N(0, \mathbf{\Sigma}_\epsilon$). Within our framework, we do not compress this to a simple relation between peak magnitude and intrinsic colour; we find that much of the residual intrinsic variation of SNe Ia SEDs does not correlate with $B-V$ colour at peak. However, we emphasise that a relation between peak magnitude and intrinsic colour, independent of light curve stretch, is present in the model.

\subsubsection{Hierarchical Dust Model}
\label{model_overview}

The focus of this analysis is to study the host galaxy dust distributions of the sample. In this work, we modify the BayeSN model discussed previously by fixing all global parameters not related to host galaxy dust extinction ($W_0$, $W_1$, $\mathbf{\Sigma}_\epsilon$). These parameters are fixed to the values for the T21 model presented in \citet{T21}, which was trained on the sample of 157 SNe from Foundation DR1 also included in this work.

The hyperparameters inferred in our hierarchical model are the mean and standard deviation of the host galaxy $R_V$ population distribution ($\mu_R$ and $\sigma_R$), the population mean extinction ($\tau_A$), which defines the scale of the exponential prior on $A_V$, and the intrinsic achromatic SN scatter ($\sigma_0$). We place a uniform hyperprior on $\mu_R$ of $\mu_R \sim U(1.2, 6)$. For $\sigma_R$ we use a half-normal hyperprior with a scale factor of 2 [$\sigma_R \sim \text{Half-}N(0, 2^2)$], following \citet{T21}\footnote{In our analysis, we find that in some cases the data does not provide strong constraint on the value of $\sigma_R$. We do consider the use of different hyperpriors on $\sigma_R$ to examine the prior dependence on our posteriors, but find that this has minimal impact on our analysis so opt to focus solely on a scale factor of 2 within this work.}. For $\tau_A$ and $\sigma_0$, we adopt half-Cauchy priors with scale parameters of 1.0 mag and 0.1 mag respectively to reflect the expected scales of these parameters, as in \citet{M20}. 

\subsubsection{Conditioning on Distance}
\label{distcond}

There are two approaches that can be taken with regard to distance in this type of hierarchical model, as discussed in \citet{TM22}. The first is to condition on distance based on redshifts and an assumed cosmology, while the second is to keep photometric distance as a free parameter and marginalise over the distance to each SN $\mu^s$ when inferring dust hyperparameters. The latter can be advantageous as it provides a cosmology-independent dust estimate, but effectively means that dust properties are inferred using colour information alone rather than colour and magnitude information. For an optical plus NIR analysis such as \citet{TM22}, this is sufficient as optical-NIR colours provide additional constraints on the dust law. However, for an optical-only analysis it is necessary to condition on redshift-based distances to obtain reasonable dust constraints. 

In this work, we condition on the redshift-distance relation such that the external constraint on $\mu^s$ is,
\begin{equation}
    \mu^s | \, z^s \sim N(\mu_{\Lambda\text{CDM}}(z^s), \sigma^{2,s}_{\text{ext}})
\end{equation}
where $z^s$ is the redshift of each SN, $\mu_{\Lambda\text{CDM}}(z)$ is the distance modulus of redshift $z$ assuming a flat $\Lambda$CDM cosmology with $\Omega_m=0.28$ and $H_0 = 73.24$ km s$^{-1}$ Mpc$^{-1}$ and $\sigma_{\text{ext}}^s$ is the uncertainty in $\mu_{\Lambda\text{CDM}}(z^s)$. This is based on propagating the uncertainty in the spectroscopic redshift $z^s$ through to an uncertainty in $\mu$. The redshift uncertainty is given by the individual uncertainty $\sigma_z^s$ added in quadrature with a peculiar velocity dispersion $\sigma_\text{pec}$. In this work, we assume $\sigma_{\text{pec}}=150$~km~s$^{-1}$ \citep{Carrick15}\footnote{We use this value for consistency with previous BayeSN analyses although note that the use of a higher value e.g. 300~km~s$^{-1}$ does not impact our conclusions regarding dust. A higher peculiar velocity dispersion trades off slightly against the inferred value of $\sigma_0$.}. All posteriors presented in this work are conditioned on this fixed cosmology.

\subsubsection{Impact of Truncated Dust $R_V$ Population Distribution}
\label{Rv_distributions}

As discussed in Section \ref{intro}, a number of previous studies have applied a population distribution to model host galaxy $R_V$ for supernovae. One approach would be to model the population as a normal distribution,
\begin{equation}
    R_V^s \sim N(\mu_R, \sigma_R^2).
\label{norm_RV}
\end{equation}
However, previous works have typically used a truncated normal distribution to ensure that $R_V$ cannot go to unphysically low values. For example, \citet{BS21} used a distribution truncated at 0.5 at the lower end. A more physically-motivated lower bound on the value of $R_V$ is 1.2, based on the Rayleigh scattering limit \citep{Draine03}. We adopt this value for this analysis, modelling the host galaxy $R_V$ distribution as
\begin{equation}
    R_V^s \sim TN(\mu_R, \sigma_R^2, 1.2, \infty),
\label{trunc_RV}
\end{equation}
where $TN(\mu, \sigma^2, a, b)$ denotes a truncated normal distribution with $\mu$ and $\sigma^2$ representing the mean and variance of a normal distribution, and $a$ and $b$ representing the upper and lower bounds of truncation applied to that normal distribution. This is the $R_V$ distribution we use throughout our analysis.

However, it is important to note that using a truncated distribution will impact inference in two ways:

\begin{enumerate}
    \item Depending on the values of $\mu_R$ and $\sigma_R$, it is important to consider them as fitting parameters rather than directly as the population mean and standard deviation, which we will refer to hereafter as $\mathbb{E}[ R_V ]$ and $\sqrt{\text{Var}[R_V]}$. Using Eq. \ref{trunc_RV} with $\mu_R=1.2$ will give a half-normal distribution; $\mathbb{E}[ R_V ]$ will be significantly different from $\mu_R$. Considering a different case, for example where $\mu_R = 3.0$ and $\sigma_R = 0.2$, the truncation can also have minimal effect -- $\mu_R$ and $\sigma_R$ will be practically identical to $\mathbb{E}[ R_V ]$ and $\sqrt{\text{Var}[R_V]}$. The relations between $\mu_R$, $\sigma_R$, $\mathbb{E}[ R_V ]$ and $\sqrt{\text{Var}[R_V]}$ are detailed in Appendix \ref{appendix:trunc_norm}. We use these relations to calculate posterior distributions on the population $\mathbb{E}[ R_V ]$ and $\sqrt{\text{Var}[R_V]}$ from samples of $\mu_R$ and $\sigma_R$ along our MCMC chains.
    \item A truncated $R_V$ distribution will impact inference of $\sigma_R$. In general, unphysically low values of $R_V$ will lead to poor quality light curve fits since they do not represent realistic extinction laws. Using a normal distribution without truncation rules out larger values of $\sigma_R$ to ensure that these low $R_V$ values are not included in the distribution. With a truncated $R_V$ distribution, these larger $\sigma_R$ values are not ruled out since low $R_V$ values are already excluded - a large $\sigma_R$ will only impact the upper end of the distribution. Overall, the use of a truncated distribution can lead to higher inferred values of $\sigma_R$, compared with an untruncated distribution.
\end{enumerate}

The extent to which both of these effects will impact the analysis depends on the values of $\mu_R$ and $\sigma_R$. If the true $R_V$ distribution is far above the lower limit, the impact will be negligible. If the true distribution overlaps significantly with this lower bound, the truncation will have a significant impact. 

\subsubsection{Assessing Quality of MCMC Chains}

We use a number of standard diagnostics to assess the quality of our MCMC chains. We initialise 4 independent chains at different points of parameter space and run them independently, and verify that the chains have mixed and converged using the $\hat{R}$ (Gelman–Rubin) statistic as well as assessing the effective sample size \citep{Gelman92, Vehtari19}. In addition, we check that our chains do not contain any divergent transitions \citep[e.g.][]{Betancourt14, Betancourt17}.

\subsection{Analysis Variations}

Throughout this analysis, we perform a number of variations of our dust inference model, which are detailed as follows.

\subsubsection{Single Dust Population}

We first consider that all SNe are drawn from the same host galaxy dust distribution, with the same priors on $R_V^s$ and $A_V^s$ used across all SNe. For this approach, we consider each of the Foundation, DES3YR and PS1MD subsamples separately as well as considering one combined sample including all SNe.

\subsubsection{Binned Populations}
\label{binned_model_overview}

Motivated by ongoing debate regarding the cause of the mass step, we also consider a variation of the model where the dust population hyperparameters are binned based on global host galaxy mass. For example, using this binned approach and assuming a truncated normal distribution, the host galaxy $R_V$ distributions become
\begin{equation}
    R_V^s \sim
    \begin{cases}
        TN(\mu_{R,\text{HM}}, \sigma_{R,\text{HM}}^2, 1.2, \infty), & \text{ if } M_*^s >  M_\text{split} \\
        TN(\mu_{R,\text{LM}}, \sigma_{R,\text{LM}}^2, 1.2, \infty), & \text{ if } M_*^s <  M_\text{split} \\
    \end{cases}
\end{equation}
where $M_*^s$ is the host stellar mass of each SN $s$ and $M_\text{split}$ is some reference stellar mass at which the split point is located. Such a split is also applied to give separate $A_V$ distributions for high- and low-mass galaxies described by different $\tau_A$ parameters. In this work, we set $M_\text{split}$ at $ 10^{10} M_\odot$.

In addition, we include parameters which represent a possible intrinsic difference between the populations of SNe Ia in each mass bin. This is done so that the model is flexible enough to allow either a mass step driven by differences in dust properties, some other intrinsic effect, or some combination thereof. We consider three separate forms for this intrinsic mass step, described as follows:

\begin{enumerate}
    \item A model including mass step parameters $\Delta M_{0,\text{HM}}$ and $\Delta M_{0,\text{LM}}$ which act as a constant achromatic shift in magnitude for each bin applied to the mean of the $\delta M^s$ distributions, such that
\begin{equation}
    \delta M^s \sim
    \begin{cases}
        N(\Delta M_{0,\text{HM}}, \sigma_{0,\text{HM}}^2), & \text{ if } M_*^s >  M_\text{split} \\
        N(\Delta M_{0,\text{LM}}, \sigma_{0,\text{LM}}^2), & \text{ if } M_*^s <  M_\text{split}. \\
    \end{cases}
\end{equation}
    \item A model including mass step parameters $\Delta W_{0,\text{HM}}$ and $\Delta W_{0,\text{LM}}$, which allow for a time- and wavelength-dependent difference in the baseline intrinsic ($\theta_1=0$) SED, independent of stretch, between SNe in different environments. Throughout our analysis, we use the term baseline to refer to the intrinsic SED with $\theta_1=0$, since for a population which has a non-zero mean value of $\theta$ this does not necessarily correspond to the population mean. In this model,
\begin{equation}
    W_0 =
    \begin{cases}
        W_0^\text{T21} + \Delta W_{0,\text{HM}} & \text{ if } M_*^s >  M_\text{split} \\
        W_0^\text{T21} + \Delta W_{0,\text{LM}} & \text{ if } M_*^s <  M_\text{split} \\
    \end{cases}
\end{equation}
    where $W_0$ describes the baseline intrinsic SED as in Equation \ref{bayesn_equation}, $W_0^\text{T21}$ represents the $W_0$ matrix inferred in \citet{T21} across SNe Ia in both high- and low-mass galaxies and $\Delta W_{0,\text{HM}}$ and $\Delta W_{0,\text{LM}}$ encode a shift in the baseline intrinsic SED in each environment. $W_0$ and $\Delta W_{0,\text{HM/LM}}$ are matrices defining a 2D cubic spline surface in wavelength and time, which warp the Hsiao template to match the baseline intrinsic colours of the training sample. The matrices are of shape $6\times6$ as detailed in \citet{T21}; there are 6 knots in phase every 10 rest-frame days between -10 and +40 days relative to B-band maximum, and 6 knots in wavelength with one at the effective wavelength of each of $griz$ bands and an additional 2 knots at the high and low end acting as an anchor at the edge of the wavelength coverage of the model. $\Delta W_{0,\text{HM/LM}}$ therefore each consist of 36 parameters. This more general model in principle also allows for the previous case of a constant, achromatic magnitude shift if that is what the data supports.
    \item A model which does not include any parameters representing an intrinsic mass step and includes splits only on dust parameters.
\end{enumerate}

The first case allows for the possibility of an intrinsic, achromatic mass step between SNe in high- and low-mass galaxies, as explored in \citet{T21, TM22}. However, this form carries with it the assumption that the baseline intrinsic colour of SNe Ia is the same between high- and low-mass environments. While this is a possibility, the most general and flexible model we explore is one that allows for the baseline intrinsic SED of SNe Ia to vary between high- and low-mass galaxies, which is the second case we explore. It is important to jointly consider intrinsic colour along with the dust distribution since dust extinction is inferred with respect to intrinsic properties.

In the first case, for $\Delta M_{0,\text{HM}}$ and $\Delta M_{0,\text{LM}}$ we assume a uniform hyperprior in the range $[-0.2, 0.2]$ as in \citet{T21} to safely capture the full range of mass steps observed in previous studies. The total achromatic mass step $\Delta M_{0}$ is then given by $\Delta M_{0,\text{HM}} - \Delta M_{0,\text{LM}}$. This must be parameterised as a separate parameter for each mass bin to avoid arbitrarily setting one bin to the mean absolute magnitude of the T21 BayeSN model.

In the second case, we assume that the shift in $W_0$ is a small variation around the overall population baseline intrinsic SED; for all of the elements that compose the $\Delta W_{0,\text{HM/LM}}$ matrix, we use the hyperprior
\begin{equation}
    \Delta W_{0,\text{HM/LM}} \sim N(\mathbf{0}, 0.1\times\bm{I}).
\end{equation}
where $\bm{I}$ is the identity matrix. The factor of 0.1 represents our expectation that this effect is small\footnote{Relaxing this expectation and allowing a wider prior does not impact our conclusions.}.

In the third case, we only include splits on dust parameters between different environments with no parameters representing an intrinsic mass step, effectively enforcing a mass step which is explained by dust properties. This is the approach used by \citet{BS21, Popovic21} and we include this case for comparison with these works and to examine the effect omitting these parameters has on dust inference.

\subsubsection{Redshift Evolution}
\label{zoverview}

The samples we consider in this analysis range in redshift from 0.015 up to 0.4, factoring in the redshift cuts applied to mitigate for selection effects. The range spanned means that these SNe provide an opportunity to test whether there is any evidence that the SN Ia host galaxy dust distribution varies with redshift. Our approach of using volume-limited samples, based on redshift cuts, to test for redshift evolution is similar to that of \citet{Nicolas21}. We apply a simple linear model for $\mu_R$, adapting Eq. \ref{trunc_RV} by setting
\begin{equation}
    \mu_R = \mu_R(z) = \eta_R \times z + \mu_{R,z0}
\end{equation}
where $z$ is redshift, $\eta_R$ is the gradient of the relation between redshift and $\mu_R$, and $\mu_{R,z0}$ is the value of $\mu_R$ at $z=0$. Within the inference, the prior on a SN at redshift $z^s$ is then $R_V^s \sim TN(\mu_R(z^s), \sigma_R^2, 1.2, \infty)$. We choose a linear model for this relation as the gradient parameter $\eta_R$ provides a simple way to assess whether the data provides evidence for evolution with redshift i.e. that $\eta_R$ is non-zero. We also apply a similar linear relation with redshift to $\tau_A$, with the value of $\tau_A$ at $z=0$ denoted by $\tau_{A,z0}$ and the gradient of the relation given by $\eta_\tau$. For $\sigma_R$, we are often only able to obtain upper limits and judge that we would not be able to constrain a $\sigma_R$ redshift relation. We elect to keep $\sigma_R$ as a fixed population parameter and only allow $\mu_R$ and $\tau_A$ to vary with redshift.

We use a joint prior on $\mu_{R,z0}$ and $\eta_R$. For $\mu_{R,z0}$ we use the same prior as applied for $\mu_R$ previously. For $\eta_R$, we apply a uniform prior,
\begin{equation}
    \eta_R | \mu_{R,z0} \sim U(1.2 - \mu_{R,z0}, 6 - \mu_{R,z0}).
\end{equation}
This enforces the condition that $\mu_R$ is restricted to the range [1.2, 6] below $z=1$, regardless of $\mu_{R,z0}$, to rule out unphysical redshift evolution in the $R_V$ distribution. For $\tau_{A,z0}$, we use the same hyperprior as for $\tau_A$ in Section \ref{model_overview}. For $\eta_\tau$, we place a wide uninformative hyperprior such that $\eta_\tau \sim U(-0.5, 0.5)$ mag, chosen to prevent the posterior distribution from approaching the prior bounds.

\section{Results and Discussion}
\label{results}

\begin{figure}
\centering
\includegraphics[width = \linewidth]{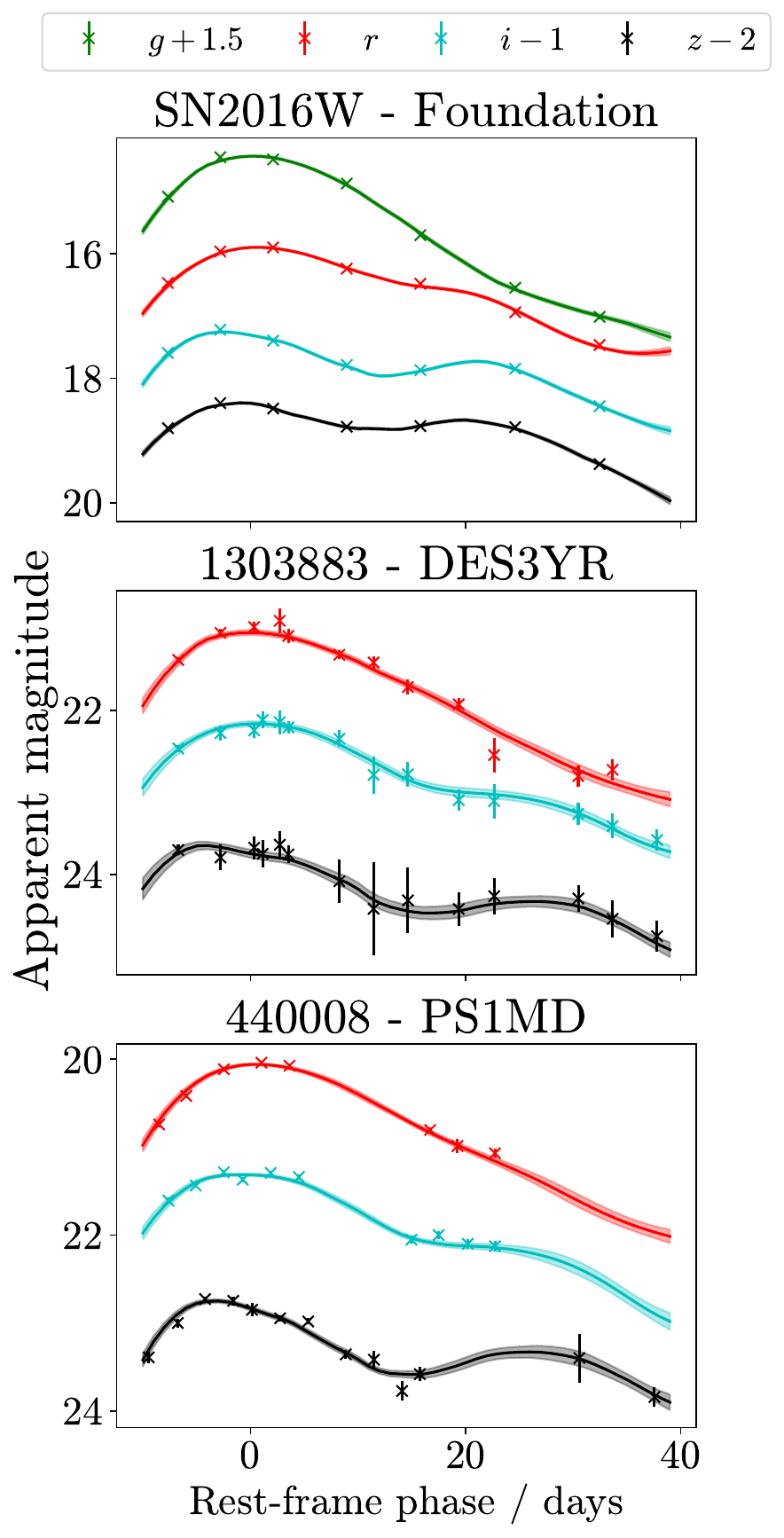}
\caption{Example BayeSN light curve fits to SN2016W from the Foundation sample, 1303883 from the DES3YR sample and 440008 from the PS1MD sample. Bands have been offset arbitrarily for clarity. Note that while the data presented is in magnitude space, all fitting is performed in flux space. Light curves are calculated for all posterior samples of SN latent parameters, and lines and shaded regions represent the mean and standard deviation of all these light curves.}
\label{example_fits}
\end{figure}

In this section we present and discuss our results for each of our model variations. To demonstrate typical light curve fits obtained using BayeSN, Figure \ref{example_fits} shows examples of a fit to a SN from each of the three surveys which make up our combined sample.

\subsection{Population Sub-sample Dust Properties}
\label{Rv_pop_results}

We begin by considering each of the DES, Foundation and PS1MD samples separately, as well as a combined sample containing all SNe across the three samples. The full results are shown in Table \ref{separate_sample_table_truncRv}.

Individually, we obtain $\mu_R$ values of $2.66\pm0.13$, $3.14\pm0.59$ and $1.69\pm0.33$ for Foundation, DES3YR and PS1MD respectively. The differences seen here perfectly highlight the importance of considering the population mean $\mathbb{E}[ R_V ]$ as well as just $\mu_R$ when using a truncated distribution. The inferred value of $\mu_R=1.69\pm0.33$ for PS1MD is close to the lower truncation bound of 1.2, and therefore a difference between $\mathbb{E}[ R_V ]$ and $\mu_R$ is expected. When we calculate $\mathbb{E}[ R_V ]$ for these samples, we instead infer $2.66\pm0.13$, $3.41\pm0.57$ and $2.44\pm0.29$ respectively for Foundation, DES3YR and PS1MD, showing PS1MD to be much more consistent with the others. While there is a numerical difference between DES3YR and PS1MD close to 1, considering the uncertainties all three of these samples are statistically consistent with each other and also with the inferred $\mathbb{E}[ R_V ]$ value of the combined population, $2.58\pm0.14$.

\begin{figure*}
\centering
\includegraphics[width = \textwidth]{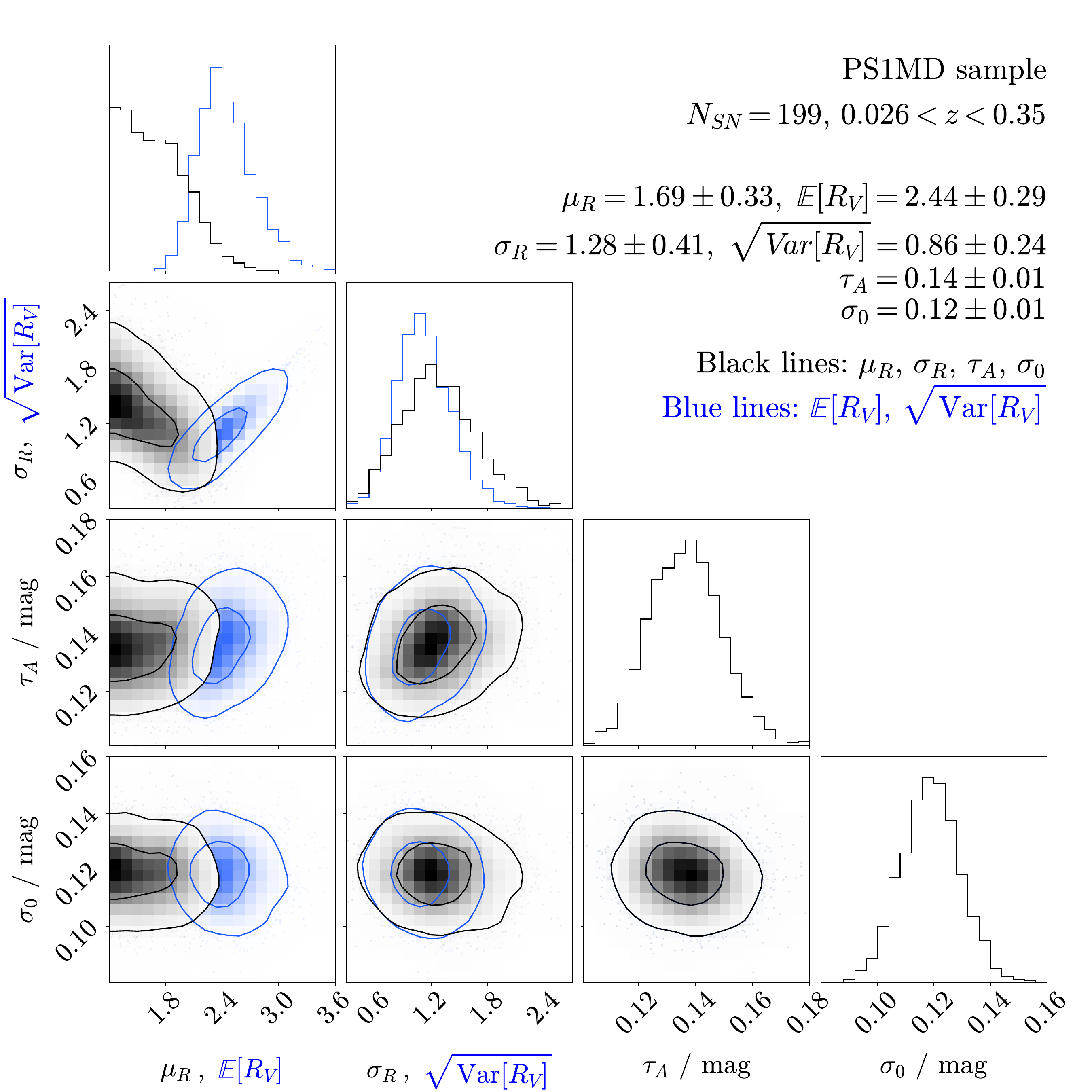}
\caption{Corner plot showing joint and marginal posteriors on $\mu_R$, $\sigma_R$, $\tau_A$ and $\sigma_0$ as well as $\mathbb{E}[ R_V ]$ and $\sqrt{\text{Var}[R_V]}$ for the PS1MD sample. This demonstrates the importance of considering $\mathbb{E}[ R_V ]$ and $\sqrt{\text{Var}[R_V]}$ rather than just $\mu_R$ and $\sigma_R$ when using truncated distributions.}
\label{PS1MD_trunc_corner}
\end{figure*}

The effect of the truncated distribution is also demonstrated in Figure \ref{PS1MD_trunc_corner}, which shows the posteriors for $\mathbb{E}[ R_V ]$ and $\sqrt{\text{Var}[R_V]}$ as well as $\mu_R$ and $\sigma_R$ for the PS1MD sample - while the $\mu_R$ posterior extends all the way to the truncation boundary at 1.2, the posterior on $\mathbb{E}[ R_V ]$ shows a clear peak well above 1.2.

Considering $\sigma_R$ and $\sqrt{\text{Var}[R_V]}$, we are able to obtain only upper limits from Foundation and DES3YR; we infer $\sigma_R=1.28\pm0.41$ and $\sqrt{\text{Var}[R_V]}=0.86\pm0.23$ from PS1MD. We obtain much greater constraint on $\sigma_R$ with the combined sample, instead inferring $\sigma_R=0.67\pm0.27$ and $\sqrt{\text{Var}[R_V]}=0.59\pm0.20$. 

For $\tau_A$, we infer $0.19\pm0.02$ mag, $0.14\pm0.02$ and $0.14\pm0.01$ mag for Foundation, DES3YR and PS1MD respectively. The higher value for Foundation is intriguing; although we have applied redshift cuts to both the DES3YR and PS1MD samples to mitigate for selection effects, it is possible that the sample is lacking higher $A_V$ objects at higher redshift which is reducing the inferred $\tau_A$ values for these samples. Nevertheless, all three individual samples are consistent with the combined sample value of $0.16\pm0.01$ mag. Finally, considering $\sigma_0$ we obtain values of $0.10\pm0.01$, $0.11\pm0.01$ and $0.12\pm0.01$ mag for Foundation, DES3YR and PS1MD respectively, and a combined sample value of $0.11\pm0.01$ mag.

\begin{table*}
\caption{Inferred parameter values when applying our hierarchical dust model to each sub-sample separately, as well as for the combined sample, assuming a truncated normal $R_V$ distribution. $\mathbb{E}[ R_V ]$ and $\sqrt{\text{Var}[R_V]}$ respectively denote the population mean and square-root of the population variance, as opposed to the fitting parameters $\mu_R$ and $\sigma_R$. Values quoted as $\mu\pm\sigma$ represent the mean and standard deviation of the posterior, while those quoted as X (Y) denote the 68th (95th) percentiles for parameters where only an upper limit could be constrained.}
\centering
\begin{tabular}{ ccccccccc }
	\hline
        Sample & $N_\text{SN}$ & $\mu_R$ & $\sigma_R$ & $\mathbb{E}[ R_V ]$ & $\sqrt{\text{Var}[R_V]}$ & $\tau_A$ / mag & $\sigma_0$ / mag \\
	\hline
        Foundation & 157 & 2.66$\pm$0.13 & 0.23 (0.47) & 2.66$\pm$0.13 & 0.23 (0.47) & 0.19$\pm$0.02 & 0.10$\pm$0.01 \\
        DES3YR & 119 & 3.14$\pm$0.59 & 1.21 (2.93) & 3.41$\pm$0.57 & 1.10 (2.04) & 0.14$\pm$0.02 & 0.11$\pm$0.01 \\
        PS1MD & 199 & 1.69$\pm$0.33 & 1.28$\pm$0.41 & 2.44$\pm$0.29 & 0.86$\pm$0.23 & 0.14$\pm$0.01 & 0.12$\pm$0.01 \\
        Combined & 475 & 2.51$\pm$0.15 & 0.65$\pm$0.27 & 2.58$\pm$0.14 & 0.59$\pm$0.20 & 0.16$\pm$0.01 & 0.11$\pm$0.01 \\
        \hline
\end{tabular}
\label{separate_sample_table_truncRv}
\end{table*}

It should be noted that our inferred parameters for the SN Ia host galaxy $R_V$ population distribution are consistent with those from previous hierarchical Bayesian analyses which incorporated both optical and NIR data. Previous analysis using BayeSN in \citet{TM22} of 75 nearby SNe from CSP-I \citep{Krisciunas17} found $\mu_R=2.59\pm0.14$ and $\sigma_R=0.62\pm0.16$, while a light curve model-independent analysis of dust reddening based on peak optical and NIR apparent colours of 65 low-redshift SNe Ia presented in \citet{Ward23} found $\mu_R=2.61^{+0.38}_{-0.35}$ and placed 68th (95th) percentile upper limits of $\sigma_R < 0.92 (1.96)$.

\subsection{Dust Properties Split by Host Mass}
\label{Rv_split_results}

We next consider the dust properties binned based on global host galaxy mass. As discussed in Section \ref{binned_model_overview}, we consider three separate variants of this model: an achromatic intrinsic mass step, a difference in baseline intrinsic SED between SNe in high- and low-mass galaxies, and a less flexible model which only incorporates differences in dust properties. The results obtained from these models are summarised in Table \ref{mass_split_table}, but discussed here in more detail.

\subsubsection{Intrinsic Achromatic Mass Step}
\label{deltaMresults}

\begin{figure*}
\centering
\includegraphics[width = \textwidth]{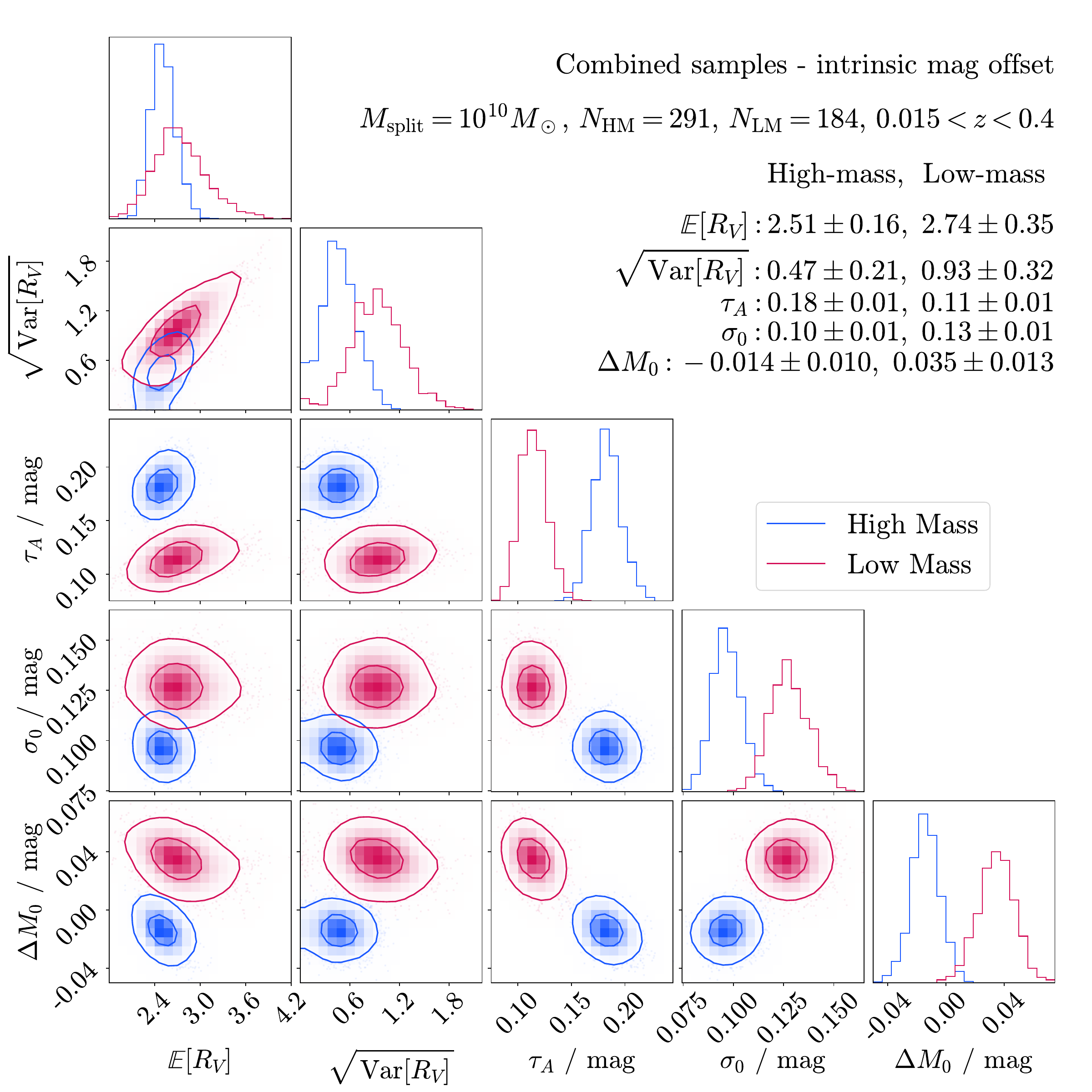}
\caption{Corner plot showing joint and marginal posteriors on $\mu_R$, $\sigma_R$, $\tau_A$, $\Delta M_0$ and $\sigma_0$ for high- and low-mass hosts for 475 SNe across Foundation, DES3YR and PS1MD, with the split between high- and low-mass hosts at $10^{10}$ M$_\odot$. These posteriors are for a model which allows for a constant intrinsic magnitude offset between each mass bin.}
\label{combined_HMvLM_notrunc}
\end{figure*}

We first consider dust properties under the assumption of an achromatic mass step, with a wavelength-independent intrinsic magnitude offset between SNe in high- and low-mass host galaxies. Figure \ref{combined_HMvLM_notrunc} shows joint and marginal posterior distributions of $\mathbb{E} [ R_V ]$, $\sqrt{\text{Var}[R_V]}$, $\tau_A$, $\Delta M_0$ and $\sigma_0$ for each mass bin (Figure \ref{combined_HMLMdiff_notrunc} in Appendix \ref{appendix:diff_plots} shows the posteriors for the differences in inferred parameter values between each mass bin).

In this case, we find evidence in favour of an intrinsic offset, inferring $\Delta M_0=-0.049\pm0.016$ at a significance of 3.1$\sigma$. Moreover, using this model we find no evidence for a difference in $R_V$ as a function of host galaxy stellar mass; for $\mathbb{E}[ R_V ]$ we infer 2.51$\pm$0.16 and 2.74$\pm$0.35 respectively across high- and low-mass host galaxies, with $\Delta \mathbb{E}[ R_V ]=0.32\pm0.58$, consistent with zero. 

In addition to the $R_V$ distributions and $\Delta M_0$, we consider $\tau_A$ and $\sigma_0$ separately for high- and low-mass hosts. $\tau_A$ and $\Delta M_0$ have similar effects in that increasing both corresponds to a fainter observed SN sample, however the key difference is that $\Delta M_0$ is achromatic while increasing $\tau_A$ will also cause the sample to appear redder. We find evidence that $\tau_A$ is higher for high-mass hosts than low-mass hosts with a difference of 0.068$\pm$0.018 mag at a significance of $\sim$3.7$\sigma$, indicating that there is on average more dust along the line of sight to SNe Ia in high-mass galaxies than low-mass galaxies. This result is consistent with expectations from analysis of galaxy attenuation \citep[e.g.][]{Salim18, Nagaraj22, Alsing24}, although it should be noted that we do not necessarily expect consistent findings when considering line-of-sight SN extinction as opposed to galaxy attenuation.

It is noticeable that the posterior distributions for the parameters relating to $R_V$ are wider for the low-mass bin than for the high-mass bin. In part, this can be simply understood as a consequence of there being more SNe in the high-mass bin. However, we can see from the inferred $\tau_A$ values that SNe in the high-mass bin have more dust along the line-of-sight on average and therefore provide greater constraint on the properties of the dust.

Finally, concerning $\sigma_0$ we infer a slightly lower value for high-mass hosts, $0.10\pm0.01$ mag as opposed to $0.12\pm0.01$ mag with a difference of -0.031$\pm$0.012 mag at a significance of $\sim$2.6$\sigma$.

\subsubsection{Intrinsic SED Difference}

We next consider a more flexible variation of the model, with the baseline ($\theta_1=0)$ intrinsic SED able to vary between high- and low-mass galaxies. As mentioned previously, any reference to the baseline intrinsic properties refers to the case with $\theta_1=0$, i.e. independent of light curve stretch.

In this case, the parameters comprising $\Delta W_0$ are less directly interpretable in a physical sense compared with a specific mass step term $\Delta M_0$. However, $W_0$ corresponds to the baseline intrinsic SED of the SN population; for each posterior sample of $\Delta W_{0,\text{HM/LM}}$ we can integrate the SED through a given set of passbands in order to calculate intrinsic baseline light and colour curves. In this analysis, we consider derived posterior distributions on mean light and colour curves for SNe in high- and low-mass galaxies separately. We do this comparison by setting $\theta_1^s=A_V^s=\epsilon^s(\lambda,t)=0$\footnote{In principle, $\epsilon(\lambda,t)$ has the flexibility to impose differences in the mean colour distribution of SNe in each mass bin independently of $W_0$. However, we verify that the posterior samples of $\epsilon(\lambda,t)$ for both SNe in high- and low-mass galaxies separately do not shift the population mean intrinsic colours.} to assess the baseline dust- and stretch-independent properties of SNe Ia in different environments, after post-processing our MCMC chains to ensure a consistent definition of $\theta_1$ between high- and low-mass galaxies as discussed in Appendix \ref{appendix:postprocessing}.

Figures \ref{HMLM_lcs} and \ref{HMLM_cs} respectively show the posterior distributions of baseline dust- and stretch-independent $griz$ light curves and $g-r$, $r-i$ and $i-z$ colour curves for SNe in high- and low-mass galaxies, along with the differences between the two. Figure \ref{combined_HMvLM_Wsplit}, meanwhile, shows joint and marginal posterior distributions of $\mathbb{E} [ R_V ]$, $\sqrt{\text{Var}[R_V]}$, $\tau_A$ and $\sigma_0$ as well as derived distributions of baseline intrinsic peak $g$-band absolute magnitude and $g-r$ colour for each mass bin (Figure \ref{combined_HMLMdiff_Wsplit} in Appendix \ref{appendix:diff_plots} shows the posteriors for the differences in inferred parameter values between each mass bin.

Compared with the $\Delta M_0$ case presented in Section \ref{deltaMresults}, we are now analysing intrinsic differences as a function of both wavelength and time. Regarding an intrinsic mass step, in the conventional Tripp formula \citep{Tripp98} approach for estimating distances to SNe Ia we expect that an intrinsic mass step corresponds to a difference in peak magnitude in some reference band. For this model we will define $\Delta M_0 = M_\text{peak,g,HM}^\text{int} - M_\text{peak,g,LM}^\text{int}$ and infer $\Delta M_0 = -0.049\pm0.027$ mag.

Compared with the achromatic mass step model, we observe a similar magnitude offset but at a lower significance of 1.8$\sigma$. Of course, in isolation a 1.8$\sigma$ offset is not significant but it is unsurprising that our uncertainties increase when switching to a more flexible model. 

Beyond a simple mass step parameter, we should consider how the difference in baseline light and colour curves varies with both time and wavelength. We can consider the intrinsic magnitude difference at peak in different bands, inferring $\Delta M_{r,\text{peak}}^\text{int}=-0.027\pm0.022$, $\Delta M_{i,\text{peak}}^\text{int}=-0.032\pm0.019$ and $\Delta M_{z,\text{peak}}^\text{int}=-0.016\pm0.030$. The relatively large uncertainties for these values makes it challenging to comment on the wavelength dependence of any magnitude offset at peak. The most significant difference in magnitude is at 20 days post-peak in $i$-band, where $\Delta M_{i,t=20}^\text{int} = -0.099\pm0.022$ at a significance of 4.5$\sigma$. Concerning colour, at peak we see a difference in baseline intrinsic $g-r$ colour between high- and low-mass of $-0.022\pm0.010$ at a significance of 2.2$\sigma$, weak evidence that SNe Ia in high-mass galaxies are bluer around peak than those in low-mass galaxies. However, we can also consider other bands and phases; 20 days post-peak, the difference in baseline intrinsic $r-i$ colour is $0.067\pm0.015$ at a significance of 4.5$\sigma$. Overall, our results do indicate statistically significant intrinsic differences between SNe Ia in high- and low-mass host galaxies.

It is interesting to note that our results suggest that intrinsic colour differences between SNe Ia in each mass bin are larger at later times and longer wavelengths, most notably around the second peak in $i$-band. Other works have suggested that the effect of dust can itself vary in time, perhaps because of circumstellar dust or nearby dust clouds \citep[e.g.][]{Forster13, Bulla18a, Bulla18b}. Compared with these analyses, we assume that dust properties are constant with time and allow for spectrotemporal perturbations around the mean intrinsic SED. The possibility that differences in intrinsic colour between different environments vary in size with time would have significant implications for studies considering time-varying dust, and should be taken into consideration in future work.

In terms of host galaxy dust properties, for the population mean of the $R_V$ distribution for high-mass bin we infer $\mathbb{E}[ R_V ] = 2.26\pm0.14$ while for the low-mass bin we infer $\mathbb{E}[ R_V ] = 3.36\pm0.51$; the difference between the values is $\Delta \mathbb{E}[ R_V ] = -1.10\pm0.53$ at a significance of 2.1$\sigma$. As in Section \ref{deltaMresults}, we see a difference in $\tau_A$ between SNe Ia in high- and low-mass hosts, although this difference is reduced to -0.047$\pm$0.024 mag at a significance of just under 2$\sigma$. Regarding $\sigma_0$, compared with the achromatic mass step model we infer the same value for the high mass bin of $0.10\pm0.01$ mag and a slightly lower value for low-mass galaxies of $0.12\pm0.01$ mag, with the difference between high- and low-mass reduced to $-0.022\pm0.012$ mag.

Beyond considering the posterior distributions of the parameters, we can also compare SNe in high- and low-mass hosts by summarising their physical properties. In Figure \ref{BS21_plot}, we present the mean and standard error on the mean in Hubble residual binned as a function of rest-frame peak $B-V$ apparent colour (derived from the posterior samples of the latent SN parameters). These values were obtained by fitting the SN sample with the BayeSN model with the values of $\mu_R$, $\sigma_R$, $\tau_A$, $W_0$, and $\sigma_0$ fixed to the medians of the posterior samples for high- and low-mass hosts separately -- for these fits, no external distance constraint based on redshift was included. Based on this result, we see no discernible trend between Hubble residual and apparent colour nor a difference in this trend between each bin, although the lack of redder SNe in the low-mass bin makes this comparison challenging.

\begin{figure*}
\centering
\includegraphics[width = \linewidth]{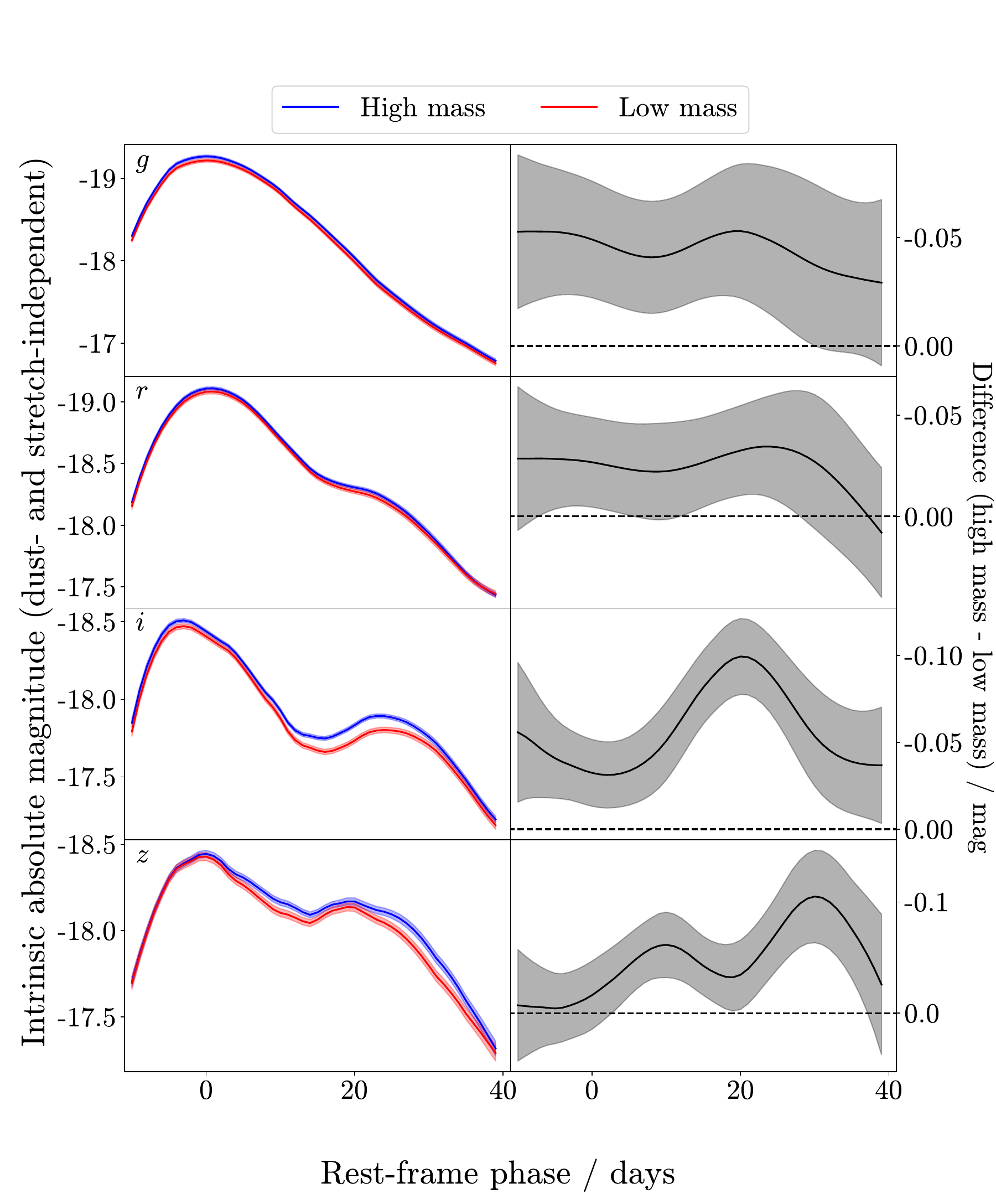}
\caption{\textbf{Left panels:} Posterior inference of the baseline dust- and stretch-independent intrinsic absolute SN Ia light curve in each of the $griz$ bands, evaluated from the posterior samples of $W_{0,\text{HM/LM}}$. Solid lines and shaded regions represent the posterior mean and standard deviation (uncertainty) of the baseline light curve in each mass bin and band. \textbf{Right panels:} Posterior inference of the difference between baseline light curves depicted in left panels. Solid lines and shaded regions again represent the posterior mean and standard deviation (uncertainty) of the difference.}
\label{HMLM_lcs}
\end{figure*}

\begin{figure*}
\centering
\includegraphics[width = \linewidth]{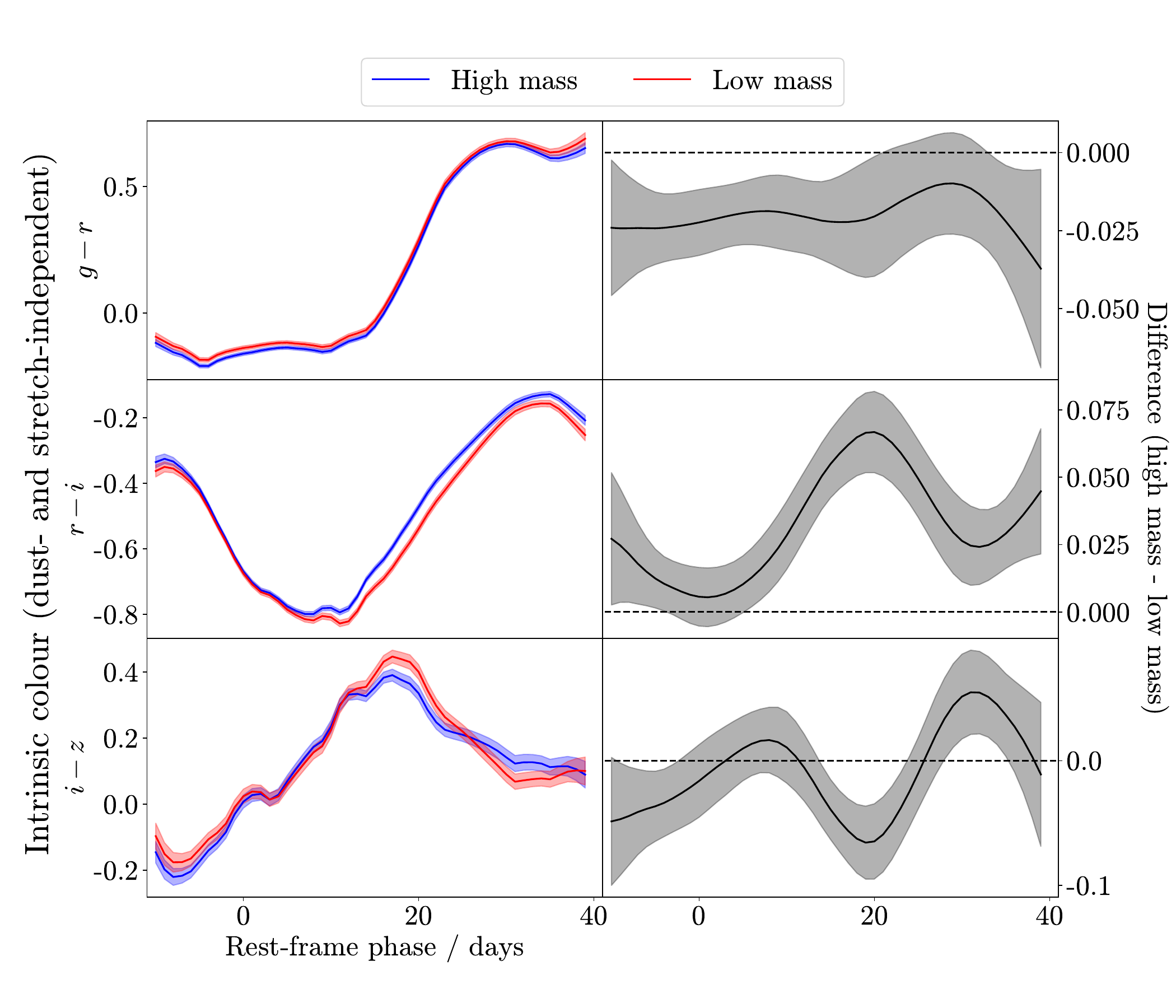}
\caption{\textbf{Left panels:} Posterior inference of the baseline dust- and stretch-independent intrinsic  SN Ia colour curve for $g-r$, $r-i$ and $i-z$ colours, evaluated from the posterior samples of $W_{0,\text{HM/LM}}$. Solid lines and shaded regions represent the posterior mean and standard deviation (uncertainty) of the baseline colour curve in each mass bin and band. \textbf{Right panels:} Posterior inference of the difference between baseline intrinsic colour curves depicted in left panels. Solid lines and shaded regions again represent the posterior mean and standard deviation (uncertainty) of the difference.}
\label{HMLM_cs}
\end{figure*}

\begin{figure*}
\centering
\includegraphics[width = \textwidth]{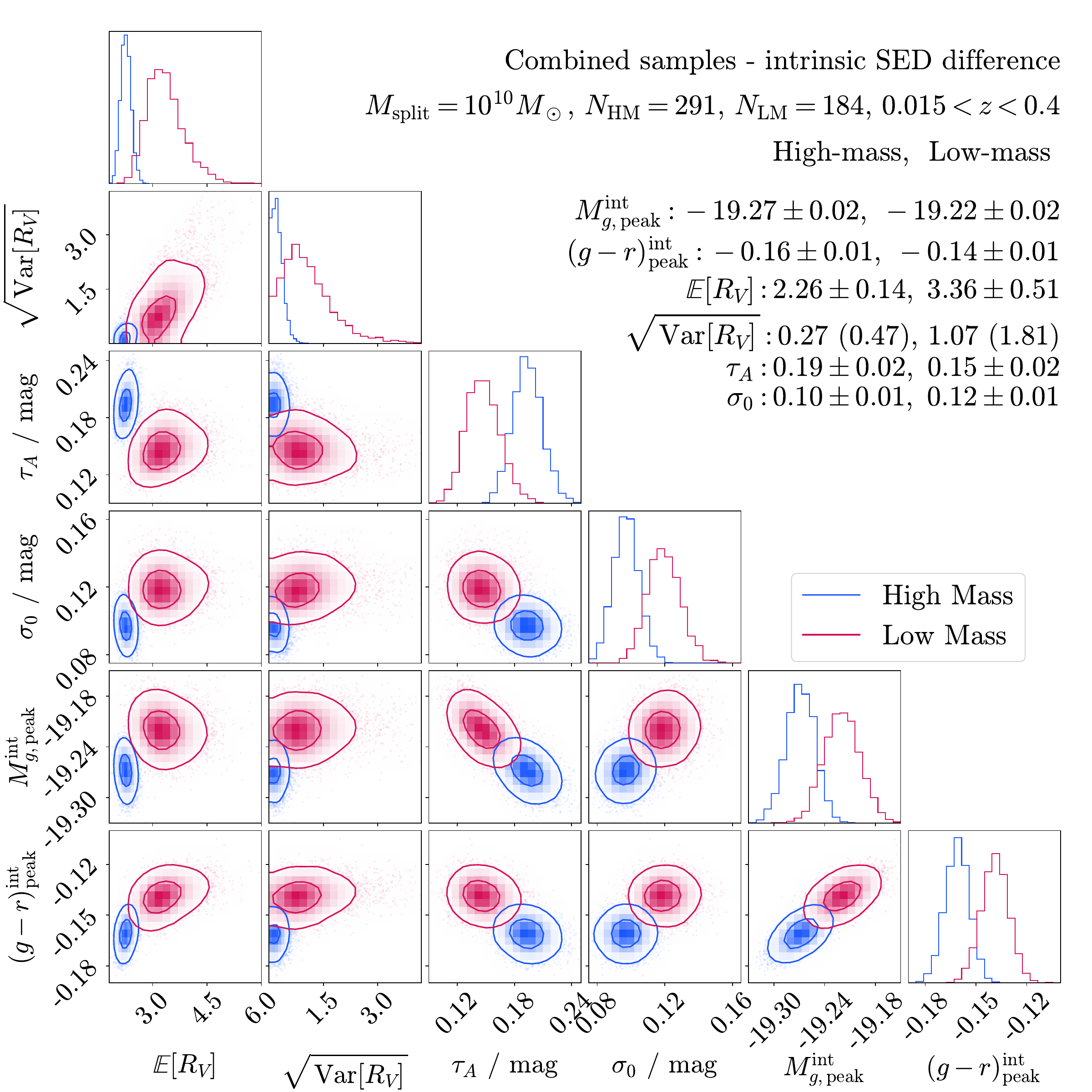}
\caption{Corner plot showing joint and marginal posteriors on $\mathbb{E} [ R_V ]$, $\sqrt{\text{Var}[R_V]}$, $\tau_A$ and $\sigma_0$, as well as derived baseline intrinsic peak $g$-band absolute magnitude and $g-r$ colour, for high- and low-mass hosts for 475 SNe across Foundation, DES3YR and PS1MD. This model allows for a difference in baseline intrinsic SED between each mass bin.}
\label{combined_HMvLM_Wsplit}
\end{figure*}

\begin{figure}
\centering
\includegraphics[width = \linewidth]{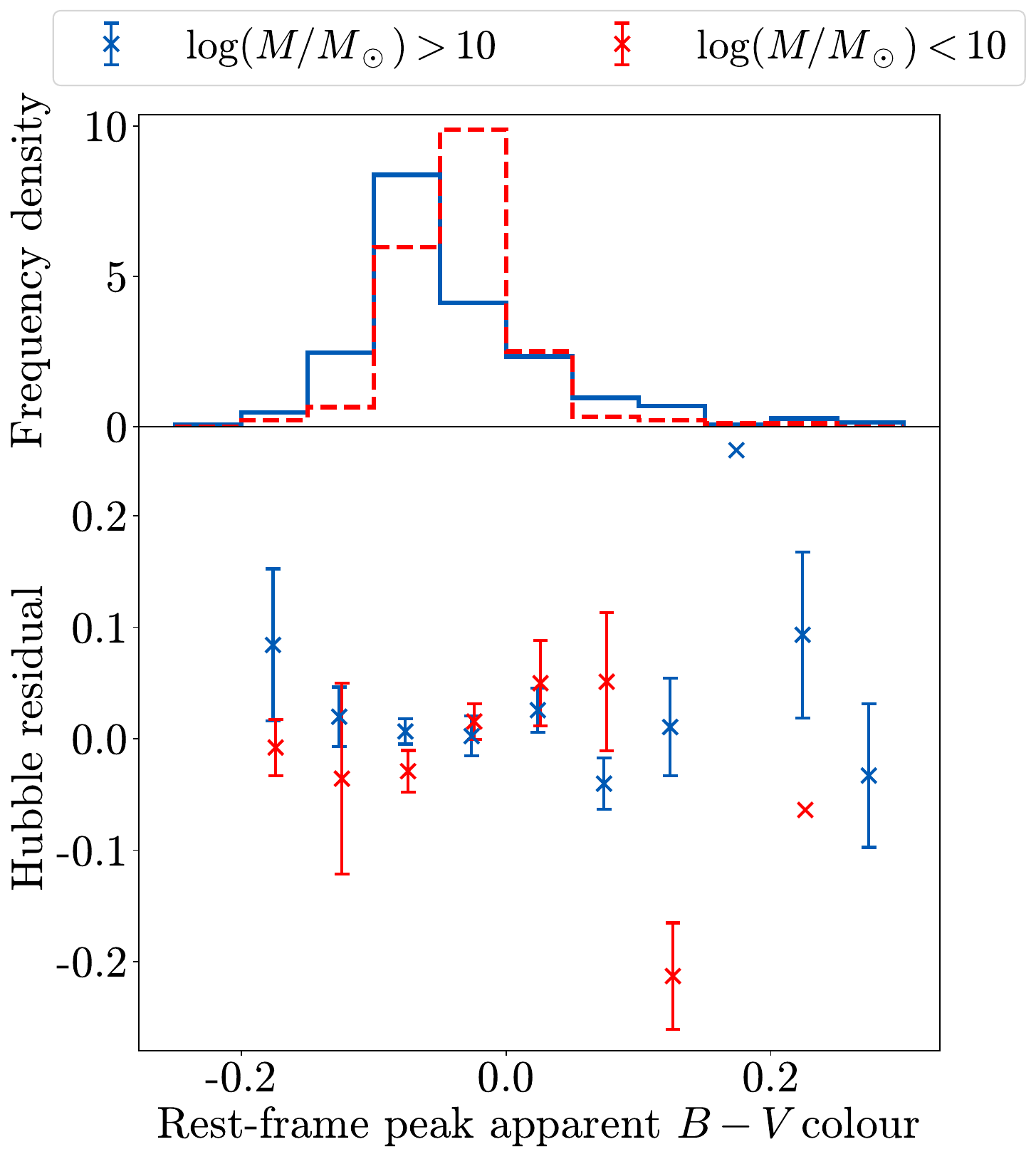}
\caption{\textbf{Upper panel:} Distributions of rest-frame apparent $B-V$ colour at peak in both high and low-mass bins. \textbf{Lower panel:} Binned Hubble residual mean and standard error on the mean obtained from fitting SNe in high- and low-mass hosts with the BayeSN model, based on the population parameters inferred when treating these samples separately, binned as a function of rest-frame apparent $B-V$ colour at peak. Bins are from -0.25 to 0.25 with widths of 0.05 -- note that the binning used was identical for high- and low-mass hosts and the bins are offset slightly for clarity. Please note that we have restricted the y-axis range for presentation purposes; there are two data points that lie outside this range which are for bins containing single objects that have large Hubble residuals.}
\label{BS21_plot}
\end{figure}

\subsubsection{Dust-only Difference}

The final case we consider is a model that does not include either of the intrinsic differences assessed previously (or, equivalently, forces these differences to be equal to zero). Under the strong assumption that the SN Ia population is intrinsically identical between high- and low-mass galaxies, for the host galaxy $R_V$ distribution we infer $\mathbb{E}[ R_V ]$ of $2.39\pm0.13$ and $3.14\pm0.39$ for high- and low-mass galaxies respectively, with a difference $\Delta\mathbb{E}[ R_V ]=-0.74\pm0.41$. As for the other models which incorporate intrinsic differences, in this case we still infer a difference in $\tau_A$ between the different mass bins; $0.18\pm0.01$ mag for high-mass and $0.13\pm0.01$ mag for low-mass, with $\Delta\tau_A=0.047\pm0.017$ mag. There is also a difference in inferred $\sigma_0$ between the two bins, with $\Delta\sigma_0=-0.029\pm0.012$ mag in this case.

\subsubsection{Intrinsic or Extrinsic?}

The aim of this binned population analysis is to investigate whether the mass step is driven by intrinsic differences, differences in the host galaxy dust properties or some combination of those two effects. However, before considering the relative contribution of each of these effects we start with a simpler question: are there intrinsic differences between SNe Ia on either side of the mass step? 

We have considered two separate model variations which jointly infer intrinsic properties in each bin simultaneously with host galaxy dust properties. The important thing to emphasise is that both of these models are flexible and allow for the possibility of a mass step driven purely by dust; nevertheless, when we condition our models on observed data the inferred parameter values support the existence of non-zero intrinsic differences. When we allow for an intrinsic achromatic magnitude offset between each mass bin, we infer a mass step $\Delta M_0=-0.049\pm0.016$ mag at a significance of 3.1$\sigma$. In the more flexible case of a difference in baseline intrinsic SED between mass bins, we infer a similar magnitude offset in $g$-band of $-0.049\pm0.027$ mag although at a lower significance of 1.8$\sigma$, unsurprising given the increased complexity of the model. When looking at intrinsic SED differences we can additionally examine differences as a function of both wavelength and time. Our analysis shows a difference in intrinsic peak $g-r$ colour between mass bins of 2.2$\sigma$ significance, providing weak evidence that SNe Ia in high-mass galaxies are intrinsically bluer than their low-mass counterparts in these bands. Looking to other wavelengths and phases, we also find much more significant differences. For example, around the time of the second maximum 20 days post-peak we find that there are differences in baseline intrinsic $i$-band magnitude and $r-i$ colour of 4.5$\sigma$ significance. Overall, our results do support the existence of intrinsic differences. 

\citet{BS21, Popovic23} posit that the mass step is driven solely by differences in dust properties between high- and low-mass galaxies. Our analysis, however, does find significant intrinsic differences and therefore contradicts this idea. Our results also differ from those of \citet{Karchev23b}, which disfavours both intrinsic and extrinsic differences between SNe Ia in different mass bins. We note that the main difference between our approach and that of \citet{Karchev23b} is that we have also treated the population mean $V$-band extinction ($\tau_A$) as a separate parameter for each mass bin. We next consider what our results can tell us about the relative contributions of intrinsic and extrinsic effects in driving the mass step.

As part of our analysis, we do consider a model which assumes no intrinsic differences between mass bins and only allows the dust properties to vary. We find a lower population mean $R_V$ of $\mathbb{E} [ R_V ]=2.39\pm0.13$ for SNe Ia in high-mass galaxies compared with low-mass galaxies, for which we infer $\mathbb{E} [ R_V ]=3.14\pm0.39$. This is similar to the trend found in \citet{BS21, Popovic23}. In this case, it is not surprising that we infer differences in the host galaxy dust properties; by definition, differences in dust properties were the only effect included in the model which were allowed to explain the mass step. 

When we increase the flexibility of our model to jointly infer an achromatic magnitude offset between mass bins alongside dust properties, our results no longer provide any evidence to support a difference in $R_V$ distribution; for high- and low-mass bins respectively, we infer 2.51$\pm$0.16 and 2.74$\pm$0.35 for $\mathbb{E} [ R_V ]$. When we allow for a magnitude offset, our results support the idea of a mass step driven solely by intrinsic differences. However, while this model is more flexible than the previous case it still carries the assumption that the baseline intrinsic colours of SNe Ia in each mass bin are identical, only allowing for a magnitude shift.

For the final case, which allows for both time and wavelength-dependent differences in the baseline intrinsic SED between each mass bin, we infer $\mathbb{E} [ R_V ]$ values of 2.26$\pm$0.14 and 3.36$\pm$0.51 for high- and low-mass bins respectively. The difference between these two values is $\Delta \mathbb{E} [ R_V ] = -1.10\pm0.53$ at a significance of 2.1$\sigma$. Our interpretation of these results is somewhat limited by the weak constraint we are able to place on $\mathbb{E} [ R_V ]$ for the low-mass bin, given the lower overall extinction of SNe Ia in this bin. Considering Hubble residuals and apparent peak $B-V$ colours, we see no evidence for a relation between the two nor any difference in this relation between each mass bin. However, this analysis is again limited by the lack of redder SNe in the low-mass bin. Our results do not provide strong evidence in favour of a mass step driven in part by extrinsic differences, but nor can we rule it out.

\citet{Gonzalez-Gaitan21} speculates that the mass step may in part result from a difference in intrinsic colour between SNe Ia in high- and low-mass bins. The Tripp formula used to estimate distance in SALT-based analyses involves a linear correction based on $B-V$ colour at peak, $\beta c$. \citet{Gonzalez-Gaitan21} therefore conjecture that a difference in intrinsic colour $c_\text{int}$ between each mass bin would lead to an overall magnitude offset $\sim\beta_\text{int}c_\text{int}$, leading to an observed mass step. We again emphasise that the BayeSN model does contain a stretch-independent colour term within $\epsilon(\lambda, t)$, which aims to capture the intrinsic variation across the SED as a function of both time and wavelength. Nevertheless, if we think of the intrinsic differences we see in terms of a linear relation this would imply $\beta_{\text{int},gr} \sim 2.2$, where $gr$ denotes the fact that this is based off $g$-band magnitude and $g-r$ colour rather than $B-V$.

The flexible model we have used, equitably treating differences in the intrinsic baseline SED of SNe Ia in each mass bin alongside differences in dust properties, has a number of advantages. However, the complexity of the effects described by the model and their relative interplay makes this a challenging problem to solve. We observe intrinsic differences in peak $g$-band magnitude and $g-r$ colour with significances of 1.8-2.2$\sigma$ (albeit with intrinsic differences of 4.5$\sigma$ at other phases), as well as a difference in $\Delta \mathbb{E} [ R_V ]$ with 2.1$\sigma$ significance. The combined sample of 475 SNe Ia used in this work has provided evidence in favour of intrinsic differences between SNe Ia in different mass bins, but further analysis of larger samples is necessary to allow for more confident conclusions about the relative contributions of intrinsic and extrinsic effects in explaining the mass step.

We have not considered different metrics for model comparison, for example using Bayes factors or the Bayesian Information Criterion, as part of this analysis. These metrics can be challenging to calculate for hierarchical models. However, we emphasise that our more complex models always allow for the possibility of the simpler cases, if that is what the data supports. For example, if the data were consistent with a mass step driven solely by differences in dust properties, we would expect to infer differences relating only to dust, rather than any intrinsic effect, using our intrinsic SED difference model. In this work we have considered a variety of models with increasing flexibility, rather than comparing a number of distinct models. The approach of considering multiple separate effects within the same model is referred to as `continuous model expansion' \citep{Gelman13}.

Within this work, we have focused on SN Ia properties split based on their host galaxy stellar mass. However, as mentioned in Section \ref{intro} previous work has established relations between SN Ia luminosity and other galaxy properties besides stellar mass, for example SFR, sSFR and rest-frame colour. These relations are also stronger when considering the properties of the local environment in which the SN exploded rather than global properties of the entire host galaxy. The BayeSN model we have used for this work can be modified to incorporate population splits on other parameters, and future studies can analyse intrinsic and extrinsic differences as a function of any host property using our approach.

\begin{table*}
\caption{Inferred parameter values for our hierarchical dust model with separate population parameters for SNe in host galaxies above and below $10^{10} M_\odot$ for the combined sample of 475 SNe Ia. We present results for cases allowing for an intrinsic magnitude difference between bins, an intrinsic SED difference and no intrinsic difference. In some cases, $\Delta$ values are presented which correspond to the difference between posterior samples for SNe in high-mass hosts and low-mass hosts.}
\centering
\begin{tabular}{ cccccccc}
	\hline
	Method & $\Delta M_0$ / mag & $\Delta (g-r)_\text{peak}^\text{int}$ & \multicolumn{2}{c}{$\mathbb{E}[ R_V ]$} & \multicolumn{2}{c}{$\sqrt{\text{Var}[R_V]}$} & ... \\
        & & &  HM & LM & HM & LM  \\
        \hline
        \hline
        Intrinsic mag diff. & -0.049$\pm$0.016 & -- & 2.51$\pm$0.16 & 2.74$\pm$0.35 & 0.47$\pm$0.21 & 0.93$\pm$0.32 \\
        Intrinsic SED diff. & -0.049$\pm$0.027 & -0.022$\pm$0.010 & 2.26$\pm$0.14 & 3.36$\pm$0.51 & 0.27 (0.49) & 1.07 (1.81) \\
        No intrinsic diff. & -- & -- & 2.39$\pm$0.13 & 3.14$\pm$0.39 & 0.43$\pm$0.20 & 1.09$\pm$0.43 \\
        \hline
\end{tabular}
\begin{tabular}{ cccccccccccc}
	\hline
	... & \multicolumn{2}{c}{$\tau_A$ / mag} & \multicolumn{2}{c}{$\sigma_0$ / mag} & $\Delta \mathbb{E}[ R_V ]$ & $\Delta \sqrt{\text{Var}[R_V]}$ & $\Delta\tau_A$ / mag & $\Delta\sigma_0$ / mag \\
         & HM & LM & HM & LM & \\
        \hline
        \hline
         & 0.18$\pm$0.01 & 0.11$\pm$0.01 & 0.10$\pm$0.01 & 0.13$\pm$0.01 & -0.23$\pm$0.38 & -0.47$\pm$0.38 & 0.068$\pm$0.018 & -0.031$\pm$0.012 \\
         & 0.19$\pm$0.02 & 0.15$\pm$0.02 & 0.10$\pm$0.01 & 0.12$\pm$0.01 & -1.10$\pm$0.53 & -0.69$\pm$0.53 & 0.047$\pm$0.024 & -0.022$\pm$0.012 \\
         & 0.18$\pm$0.01 & 0.13$\pm$0.01 & 0.10$\pm$0.01 & 0.13$\pm$0.01 & -0.74$\pm$0.41 & -0.66$\pm$0.48 & 0.047$\pm$0.017 & -0.029$\pm$0.012  \\
        \hline
\end{tabular}
\label{mass_split_table}
\end{table*}

\subsubsection{Potential Cause of Intrinsic Differences}

Both \citet{Jones23} and \citet{Taylor24} previously examined the mass step using SALT; the former incorporated an SED surface to capture the variation in the apparent SN SED as a function of host galaxy stellar mass, while the latter applied the standard SALT model to separate training samples of SNe in high- and low-mass hosts. These works found opposite conclusions, albeit with different methodologies and samples; \citet{Jones23} found that the SEDs of SNe Ia in high-mass galaxies were bluer than those in low-mass galaxies, while \citet{Taylor24} found the inverse. However, neither of these works aims to disentangle intrinsic colour from dust reddening, making a direct comparison with our results difficult. We find that SNe Ia in high-mass bins are on average bluer than their low-mass counterparts, but also that they typically have more dust along the line-of-sight as demonstrated by the differences in inferred $\tau_A$ values.

Given that our results have indicated intrinsic differences between SNe Ia in high- and low-mass galaxies, it is important to consider what could cause such differences. It has previously been suggested that the luminosity distribution of SNe Ia may be explained in part by differences in metallicities of the progenitor stars affecting the mass of $^{56}$Ni produced \citep[e.g.][]{Timmes03}. \citet[][see Fig. 4]{Kasen09} predicts that metallicity is expected to influence both the luminosity and decline rate of SNe Ia; for an otherwise unchanged progenitor star, an increase in metallicity will decrease the production of $^{56}$Ni, causing a fainter SN which declines more quickly. While higher metallicity will lead to an intrinsically fainter SN population overall, the relative impact of metallicity on both luminosity and decline rate means that we expect SNe Ia with higher progenitor metallicities to be intrinsically brighter, for a given value of stretch.

Our results indicate that SNe Ia in high-mass galaxies are intrinsically brighter than those in low-mass galaxies, independently of light curve stretch. High-mass galaxies will also on average have higher metallicities \citep[e.g.][]{Tremonti04}. Our findings are therefore consistent with the suggestion from \citet{Kasen09} that SNe Ia in higher metallicity environments will be brighter than those of equivalent light curve stretch in lower metallicity environments.

Additionally, previous analysis of the ejecta velocities of SNe Ia around peak in the Si II 6355$\AA$ line have provided evidence in favour of two separate populations of SNe Ia \citep{Wang09, Wang13}, consisting of `normal-velocity' SNe Ia and `high-velocity' SNe Ia. While we cannot comment on ejecta velocity in this work, given that the implementation of BayeSN is applied only to photometry\footnote{It is our plan to apply BayeSN to spectra in future, which would allow for this.}, we can consider if the intrinsic differences we see might relate to this idea of normal-velocity and high-velocity SNe Ia.

Some previous studies have indicated that ejecta velocities of SNe Ia correlate with apparent and intrinsic SN colour (e.g. \citealp{Foley11a, Foley11b, Foley12, Mandel14}, although see \citealp{Dettman21} for a counterexample based on the Foundation survey); specifically, these works suggest that high-velocity SNe are intrinsically redder than their normal-velocity counterparts. Additionally, \citet{Siebert20} finds that composite spectra of SNe Ia with negative Hubble residuals -- those that are brighter -- have higher ejecta velocities than those constructed from SNe Ia with positive Hubble residuals at phases around peak. Further analysis has found that these two SN Ia populations demonstrate an environmental dependence; while normal-velocity SNe occur across a full range of hosts, high-velocity SNe occur specifically in high-mass, high-metallicity environments \citep{Wang13, Pan15, Pan20}. \citet{Wang13} postulated that high-velocity SNe Ia are associated with young progenitor stars, though \citet{Pan20} found that these objects were not associated with particularly young environments and concluded that metallicity was the main driver behind these objects. On this basis, one might expect to observe that SNe Ia in the high-mass bin are on average intrinsically redder than those in the low-mass bin. However, our analysis finds the opposite - SNe Ia in the high-mass bin seem to be intrinsically bluer. It is important to note that our findings do not contradict previous results regarding the environmental dependence and nature of high-velocity SNe Ia. High-velocity SNe only comprise a subset of the population, and it may well be the case that there is an environmental dependence in the population of normal-velocity SNe Ia which acts to counter the colour difference expected from high-velocity SNe Ia. However, this does mean that the observed association of high-velocity SNe Ia with high-mass galaxies cannot explain the trends found in our analysis.

One proposed explanation for the mass step is of two different populations of SNe Ia originating from different progenitor systems. These are typically thought of being divided into young/prompt SNe Ia resulting from single degenerate systems comprising a white dwarf and massive companion star, and old/delayed SNe Ia resulting from double degenerate systems comprising two white dwarfs \citep[e.g.][]{Mannucci06, Rigault20, Nicolas21}. In this case, the mass step arises as a result of differences in the ages of the progenitor stars between each mass bin; more massive host galaxies are associated in general with older progenitor stars. The size of any step therefore increases with significance when considering environmental properties that more closely depend on progenitor age \citep[e.g. $U-R$ colour, local sSFR;][]{Rigault20, Kelsey21}.

The intrinsic differences we find in this work may be the result of two different populations of SNe Ia originating from progenitor systems of different ages. However, the physical nature of SN Ia explosions and how they relate to their progenitors is not yet established enough to allow us to comment on whether our results regarding intrinsic differences in luminosity and colour are consistent with a split between young and old progenitor systems.

\subsection{Redshift Evolution}
\label{Rv_z_results}

We next consider the possibility of a dust distribution which varies as a function of redshift, using the linear model as outlined in Section \ref{zoverview}. Previous work considering redshift evolution of SNe Ia host galaxy dust properties is scarce. \citet{Nordin08} considered how such an evolution might lead to systematics in inferred cosmology but did not try to infer how these properties might evolve. Recently, \citet{Thorp24} used the BayeSN model to compare dust properties between a low-redshift sample of CSP SNe Ia and a higher redshift sample from RAISIN \citep[SN IA in the IR;][]{Jones22b}, and constrained the size of the shift in $\mu_R$ between the two to be $-1.16<\Delta\mu_R<1.38$ at 95 per cent posterior probability. Although there is literature concerning relations between galaxy dust properties and other galaxy properties such as SFR and sSFR which are known to evolve with redshift \citep[e.g.][]{Salim18, Nagaraj22, Alsing24}, we note that these are based on galaxy attenuation and will not necessarily be the same when considering SN line-of-sight extinction. As such, we have no strong prior expectation on what we expect from this analysis. Our results are shown in Table \ref{Rv_z_table}.

For the gradient of the $\tau_A$ redshift relation, we infer $\eta_\tau=-0.22\pm0.06$ mag, non-zero at $\sim$3.7$\sigma$ significance. A challenge in interpreting this result is that Malmquist bias would be expected to lead to smaller inferred values of $\tau_A$ at higher redshift. While we have mitigated for this by applying selection effects to the DES3YR and PS1MD samples, some small effect may remain - overall, we caution that this result should be interpreted as an evolution in the effective $\tau_A$ in the samples we are looking at with redshift, rather than necessarily a physical evolution with redshift in the distribution of $A_V$. Nevertheless, it is important that $\tau_A$ is free to evolve with redshift in the model in order to fairly assess any evolution with redshift of $\mu_R$.

Considering the $R_V$ distribution itself, our results do not provide any evidence that it evolves with redshift across the Foundation, DES3YR and PS1MD samples - we find the gradient of the $\mu_R$-redshift relation to be $\eta_R=-0.38\pm0.70$, consistent with zero. While we have performed this analysis with $\tau_A$ also free to evolve with redshift, it should be noted that our conclusions are unchanged when we fix $\tau_A$ to be constant with redshift. However, we emphasise that the samples included in this work are all below $z=0.4$ and that further analyses with higher redshift objects are required to investigate the possibility of redshift evolution further.

\begin{table}
\caption{Inferred parameter values for our hierarchical dust model which allows $\mu_R$ and $\tau_A$ to vary with redshift.}
\centering
\begin{tabular}{ cc }
	\hline
        Parameter & Value \\
        \hline
        $\mu_{R,z0}$ & 2.58$\pm$0.17 \\
        $\eta_R$ & -0.38$\pm$0.70 \\ 
        $\tau_{A,z0}$ / mag & 0.19$\pm$0.02 \\ 
        $\eta_\tau$ / mag & -0.22$\pm$0.06 \\
        $\sigma_R$ & 0.61$\pm$0.24 \\
        $\sigma_0$ / mag & 0.11$\pm$0.01\\
        \hline
\end{tabular}
\label{Rv_z_table}
\end{table}

\section{Conclusions and Future Work}
\label{conclusion}

In this work we apply BayeSN, a hierarchical Bayesian SED model for SNe Ia, to infer the global dust distributions for samples from Foundation, DES3YR and PS1MD. We have investigated the possibility that the host galaxy mass step can be explained solely by differences in the dust population between high- and low-mass galaxies, and also perform the first hierarchical analysis allowing for a redshift evolution in the dust population. Our findings can be summarised as follows:

\begin{itemize}
    \item For a combined sample of 475 SNe Ia from Foundation, DES3YR and PS1MD (with redshift cuts applied to mitigate for selection effects), we infer the population mean of the line-of-sight $R_V$ distribution to be $\mathbb{E}[ R_V ]=2.58\pm0.14$, the square root of the $R_V$ population variance to be $\sqrt{\text{Var}[R_V]}=0.59\pm0.20$, the population mean $A_V$ to be $\tau_A=0.16\pm0.01$ mag and the intrinsic dispersion to be $\sigma_0=0.11\pm0.01$ mag. Considering each sub-sample separately, we infer values consistent with these.
    \item When jointly inferring differences in intrinsic properties simultaneously with differences in host galaxy properties, we find evidence for intrinsic differences between SNe in host galaxies with stellar masses above and below $10^{10}$ M$_\odot$. Allowing for a constant, achromatic magnitude offset between each mass bin, we infer an intrinsic mass step of -0.049$\pm$0.016 mag and no evidence for any difference in the $R_V$ distributions. However, we also apply a more flexible model which allows for time- and wavelength-dependent differences in the intrinsic SED between the two populations. In this case, we infer differences in peak stretch and dust-independent intrinsic $g$-band magnitude and $g-r$ colour of -0.049$\pm$0.027 and -0.022$\pm$0.010, and more significant differences in intrinsic $i$-band magnitude and $r-i$ colour 20 days after peak of -0.099$\pm$0.022 and 0.067$\pm$0.015. For this model we also infer a difference in the population mean of the $R_V$ distribution for each mass bin of $\Delta \mathbb{E}[ R_V ]=-1.10\pm0.53$. These results suggest that extrinsic effects, in addition to intrinsic effects, may contribute to the host galaxy mass step. Future analyses of this type on larger samples will allow for better constraints of each of these effects to better understand their relative contributions.
    \item Across all model variants we infer consistently larger line-of-sight population mean $A_V$ ($\tau_A$) values for SNe Ia in high-mass galaxies than for SNe Ia in low-mass galaxies. When allowing for a constant intrinsic magnitude offset between each mass bin we infer a difference of $\Delta\tau_A=0.068\pm0.018$ mag, although this is reduced to $\Delta\tau_A=0.047\pm0.024$ mag when allowing the baseline, stretch-independent intrinsic SED to vary with both time and wavelength between each mass bin.
    \item We apply a model which allows $\mu_R$ and $\tau_A$ to vary with redshift, modelling these relations as a straight line and constraining both the value at $z=0$ and the gradient of the relation with redshift ($\eta_R$ and $\eta_\tau$). We find no evidence that $\mu_R$ evolves with redshift over the range covered by the samples used for this work, inferring $\eta_R=-0.38\pm0.70$, consistent with 0. Concerning $\tau_A$, we infer $\eta_\tau=-0.22\pm0.06$ mag, however we emphasise that some small selection effects may remain in spite of our redshift cuts and that this result should be interpreted as an evolution in the effective $\tau_A$ in these samples rather than necessarily suggesting a physical evolution with redshift in the distribution of $A_V$. Future studies involving higher redshift objects should investigate this further.
\end{itemize}

Our flexible BayeSN, SED-based approach to studying intrinsic and extrinsic properties of SNe Ia provides an ideal framework for studying how the population varies with different host galaxy properties. As discussed in Section \ref{numpyro}, this analysis was performed using a new, GPU-accelerated implementation of BayeSN; this code has improved performance by a factor of $\sim$100, making it suitable for application to large SN samples, and is made publicly available. Further performance increases (up to an additional factor of $\sim$10) are possible through the use of Variational Inference \citep[VI;][]{Blei16}, which is planned for inclusion within BayeSN \citep{Uzsoy22}.

Alternatively, it is also possible to use the BayeSN model in an SBI framework. \citet{Karchev24} presents dust inference using BayeSN with truncated marginal neural ratio estimation (TMNRE; \citealp{Miller21} for the method itself, \citealp{Karchev23a} for a specific application to SN cosmology), an SBI approach which uses neural networks to derive posterior distributions where an analytic likelihood is not possible. Such an approach has the potential to fully incorporate survey selection effects -- which can be simulated but not expressed analytically -- within the hierarchical framework rather than requiring a redshift cut as we have used for this analysis.

In this work, we have focused on physical properties of the population of SNe Ia, conditioned on a fixed cosmology. However, the BayeSN code we make available can also be used to infer cosmology-independent distances whilst marginalising over the intrinsic and extrinsic variations in the population, and is being integrated within SNANA. Additionally, the SBI approach to dust inference using BayeSN presented in \citet{Karchev24} provides a path towards a fully hierarchical cosmological analysis of SNe Ia all the way from light curves to cosmological parameters, taking survey selection effects into account. Moving forward, BayeSN can therefore be a key component of cosmological analyses.

\section*{Acknowledgements}

MG and KSM are supported by the European Union’s Horizon 2020 research and innovation programme under ERC Grant Agreement No. 101002652 and Marie Skłodowska-Curie Grant Agreement No. 873089. ST was supported by the European Research Council (ERC) under the European Union's Horizon 2020 research and innovation programme (grant agreement no.\ 101018897 CosmicExplorer). SD acknowledges support from the Marie Curie Individual Fellowship under grant ID 890695 and a Junior Research Fellowship at Lucy Cavendish College. BMB is supported by the Cambridge Centre for Doctoral Training in Data-Intensive Science funded by the UK Science and Technology Facilities Council (STFC). EEH is supported by a Gates Cambridge Scholarship (\#OPP1144). SMW is supported by the UK Science and Technology Facilities Council (STFC).

This work was performed using resources provided by the Cambridge Service for Data Driven Discovery (CSD3) operated by the University of Cambridge Research Computing Service (\url{www.csd3.cam.ac.uk}), provided by Dell EMC and Intel using Tier-2 funding from the Engineering and Physical Sciences Research Council (capital grant EP/T022159/1), and DiRAC funding from the Science and Technology Facilities Council (\url{www.dirac.ac.uk}).

\section*{Data Availability Statement}

The data underlying this article are sourced from the Pantheon+ compilation of SNe Ia \citep{PantheonPlus}, available at \url{https://github.com/PantheonPlusSH0ES/DataRelease}. The code used for this analysis is available at \url{https://github.com/bayesn/bayesn}.




\bibliographystyle{mnras}
\bibliography{refs} 


\appendix

\section{Assessing Completeness}
\label{selection_effects}

To mitigate for the impact of selection effects such as Malmquist bias on our analysis, we have applied upper redshift cuts to the PS1MD and DES3YR samples at $z<0.35$ and $z<0.4$ respectively. Considering our model, selection effects are most likely to impact inference of the population mean $A_V$, $\tau_A$ - at higher redshifts, higher extinction objects with large $A_V$ values are less likely to be observed as they will be fainter in the observer-frame. Without mitigating for selection effects, lower values of $\tau_A$ will be inferred.

To assess our redshift cuts, we re-run our hierarchical model on the Foundation, DES3YR and PS1MD samples with lower redshift cuts to examine the impact on the results. The results of this analysis are shown in Table \ref{tau_vs_z_table}. For the DES3YR sample, lowering the upper redshift limit from 0.4 to 0.35 or 0.3 does not impact the inferred value of $\tau_A$. This provides reassurance that with our chosen redshift limit of 0.4, the sample is not significantly impacted by selection effects. Our findings are similar for the PS1MD sample; a redshift cut at 0.3 does not impact the inferred $\tau_A$ value, and while an even lower cut at 0.25 does increase the inferred value by $\sim$0.01 mag this difference is not statistically significant. Regarding Foundation, this sample focused on targets at $z<0.08$ with minimal impact from Malmquist bias \citep{Foley18}. Nevertheless, we do consider lower redshift cuts to see how this affects $\tau_A$. We find that $\tau_A$ does increase with lower redshift cuts, however these changes are only of $\sim1\sigma$ significance. We ultimately opt to use the same Foundation sample as in \citet{T21} for consistency and because the sample was gathered with minimising Malmquist bias in mind. In addition, we did explore repeating the analysis presented in this paper with lower redshift cuts for the Foundation sample, but found that it did not impact our conclusions.

\begin{table}
\caption{Inferred $\tau_A$ values using our hierarchical dust model for the Foundation, DES3YR and PS1MD samples with different upper redshift cuts.}
\centering
\begin{tabular}{ ccccc }
	\hline
        Survey & Upper redshift limit & $N_{SN}$ & $\tau_A$ / mag \\
        \hline
        Foundation & 0.06 & 123 & $0.224\pm0.021$ \\
        -- & 0.07 & 137 & $0.213\pm0.019$ \\
        -- & 0.08 & 157 & $0.194\pm0.017$ \\ 
        DES3YR & 0.3 & 65 & $0.138\pm0.022$ \\
        -- & 0.35 & 98 & $0.140\pm0.018$ \\
        -- & 0.40 & 119 & $0.142\pm0.017$ \\ 
        PS1MD & 0.25 & 108 & $0.148\pm0.017$ \\
        -- & 0.3 & 147 & $0.133\pm0.015$ \\
        -- & 0.35 & 198 & $0.137\pm0.013$ \\   
        \hline
\end{tabular}
\label{tau_vs_z_table}
\end{table}

\begin{figure}
\centering
\includegraphics[width = \linewidth]{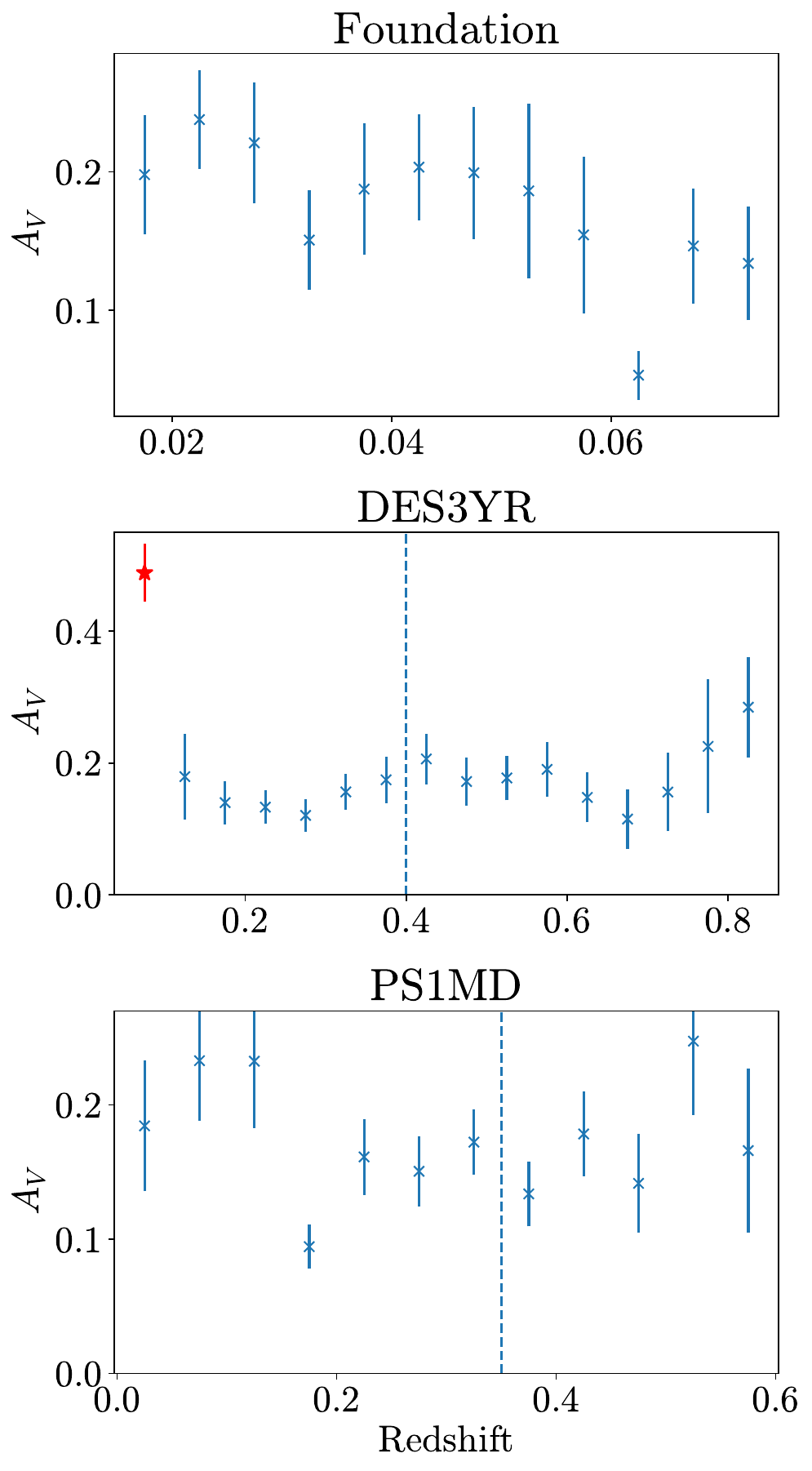}
\caption{Redshift-binned $A_V$ posterior sample mean and standard error on the mean based on BayeSN fits using the model trained in \citet{T21} for each of the Foundation, DES3YR and Foundation samples. Dashed lines for DES3YR and PS1MD indicate the upper redshift limit of the volume limited sub-samples included in our analysis. The lowest redshift bin for DES3YR, denoted by a star, includes only a single object and therefore we plot the posterior standard deviation for that object rather than a standard error on the mean.}
\label{AV_vs_z_plot}
\end{figure}

To further verify that our redshift cuts are reasonable, we also consider the inferred $A_V$ values when fitting these samples with the BayeSN model trained in \citet{T21} as a function of redshift. These results are shown in Figure \ref{AV_vs_z_plot}. For each redshift bin, we combine all posterior samples on $A_V$ across all SNe in the bin, calculate the mean and standard error on the mean for each posterior sample, and then calculate the mean across all posterior samples for these two statistics; these are the plotted values and uncertainties (with the exception of the lowest redshift bin for DES3YR, denoted by a star, in which there is only one SN and hence the plotted error bar corresponds to the posterior standard deviation on that single value.). It should be noted that many of these posteriors will only be upper limits meaning that a point estimate may not be the most appropriate summary statistic, but the mean in each bin remains useful as a first order tool to examine how $A_V$ varies with redshift. Considering the DES3YR sample, shown in the middle panel of Figure \ref{AV_vs_z_plot}, we do not see evidence for a clear decrease in $A_V$ with redshift below $z=0.4$; as mentioned previously, the single, higher $A_V$ bin denoted by a star at the lowest redshift corresponds to a single object. For the PS1MD sample, shown in lower panel of Figure \ref{AV_vs_z_plot}, $A_V$ remains consistent with redshift between  $0.15 < z < 0.35$. While some of the lowest redshift bins do have higher $A_V$ values they also contain fewer SNe.

Overall, we are confident that our samples are not significantly impacted by selection effects, although we cannot rule out the possibility that some small effect may remain.

\section{Mean and Variance of Truncated Normal Distributions}
\label{appendix:trunc_norm}

As discussed in Section \ref{Rv_distributions}, a truncated normal distribution $x \sim TN(\mu, \sigma^2, a, b)$ will not in fact have a population mean and variance $\mu$ and $\sigma^2$ - these represent the mean and variance of a normal distribution prior to truncation between $a$ and $b$. Instead, the population mean of a truncated normal distribution is given by,
\begin{equation}
    \mathbb{E}[ x ] = \mu + \frac{\phi(\alpha) - \phi(\beta)}{\Phi(\beta) - \Phi(\alpha)} \sigma
\end{equation}
while the variance is given by,
\begin{equation}
    \text{Var}[x] = \sigma^2 \Bigg[1 - \frac{\beta\phi(\beta)-\alpha\phi(\alpha)}{\Phi(\beta) - \Phi(\alpha)} - \Bigg(\frac{\phi(\alpha)-\phi(\beta)}{\Phi(\beta) - \Phi(\alpha)}\Bigg)^2\Bigg].
\end{equation}

In these definitions, $\alpha \equiv \frac{a - \mu}{\sigma}$ and $\beta \equiv \frac{b - \mu}{\sigma}$. The function $\phi(y)$ is the probability density function of the standard normal distribution,
\begin{equation}
    \phi(y) = \frac{1}{\sqrt{2\pi}}\exp\Bigg(-\frac{y^2}{2}\Bigg)
\end{equation}
and the function $\Phi(y)$ is the cumulative density function of the standard normal distribution,
\begin{equation}
    \Phi(y) = \frac{1}{2}\big(1 + \text{erf}(y/\sqrt{2})\big)
\end{equation}
where $\text{erf}(z)$ is the error function. 

In this work, we set $b = \infty$, meaning that $\phi(\beta) = 0$ and $\Phi(\beta) = 1$.  Properties of the truncated normal distribution can be found at: \url{https://people.sc.fsu.edu/~jburkardt/presentations/truncated_normal.pdf} \citep{trunc_norm}.

\section{Posterior Distributions on Differences Between Mass Bins}
\label{appendix:diff_plots}

\begin{figure*}
\centering
\includegraphics[width = \textwidth]{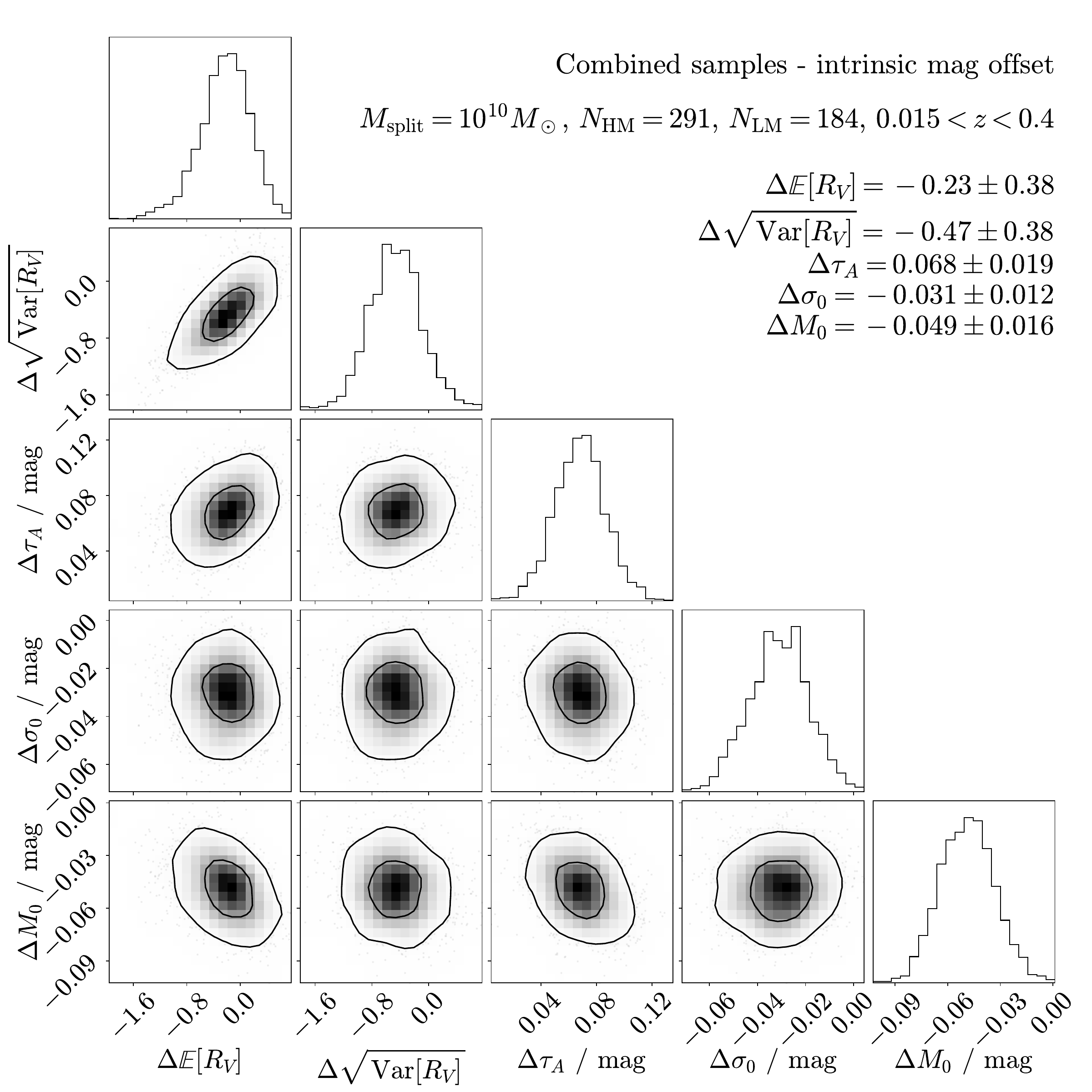}
\caption{Joint and marginal posterior distributions for the difference between inferred parameter values of $\mu_R$, $\sigma_R$, $\tau_A$, $\Delta M_0$ and $\sigma_0$ in high- and low-mass bins. These are for the model which allows for an intrinsic magnitude offset between each mass bin in addition to different host galaxy dust distributions.}
\label{combined_HMLMdiff_notrunc}
\end{figure*}

\begin{figure*}
\centering
\includegraphics[width = \textwidth]{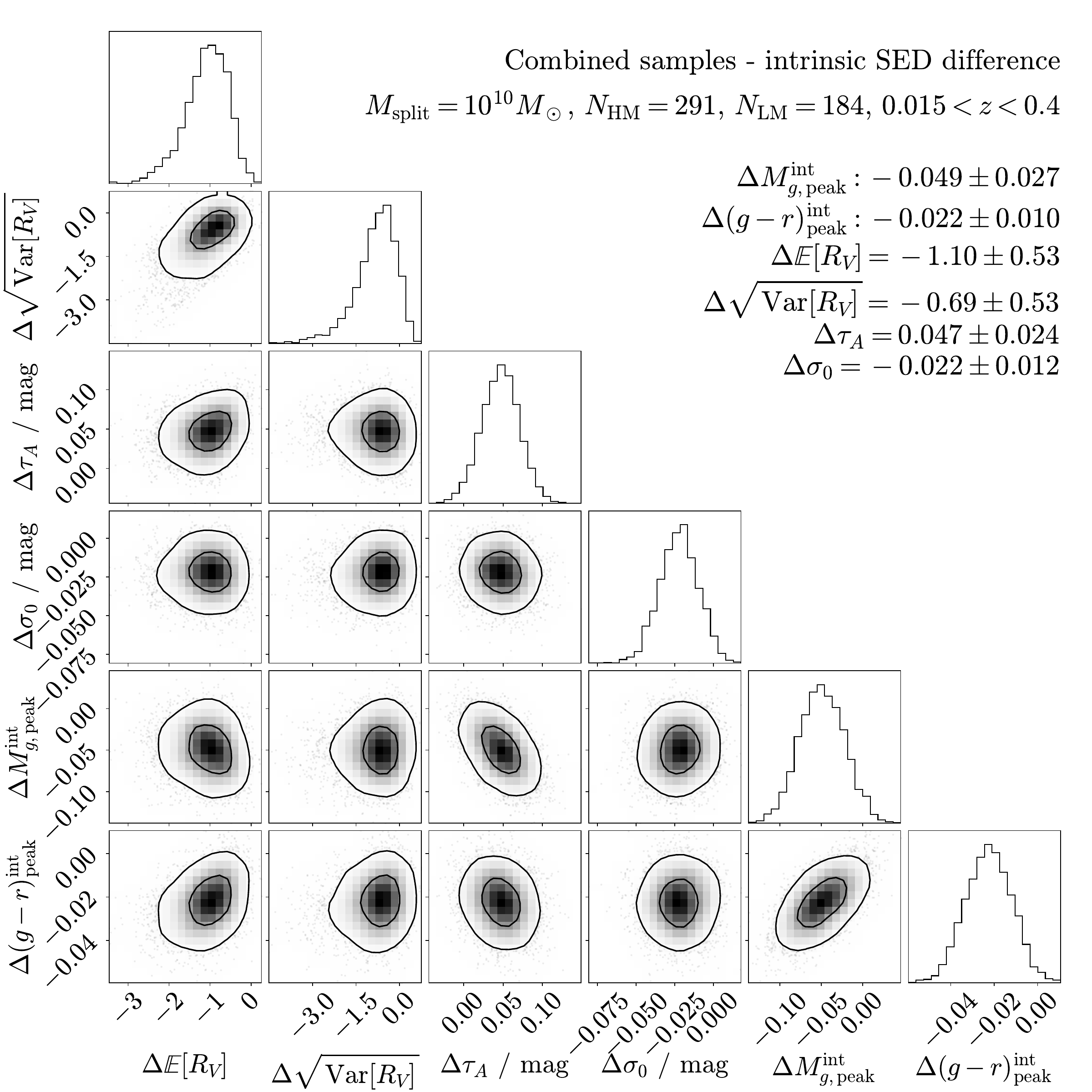}
\caption{Joint and marginal posterior distributions for the difference between inferred parameter values of $\mathbb{E} [ R_V ]$, $\sqrt{\text{Var}[R_V]}$, $\tau_A$ and $\sigma_0$, as well as derived baseline intrinsic peak $g$-band absolute magnitude and $g-r$ colour, in high- and low-mass bins. These are for the model which allows for a difference in baseline intrinsic SED between each mass bin in addition to different host galaxy dust distributions.}
\label{combined_HMLMdiff_Wsplit}
\end{figure*}

Figures \ref{combined_HMLMdiff_notrunc} and \ref{combined_HMLMdiff_Wsplit} present joint and marginal posterior distributions on the differences between the parameters inferred for each mass bin for our analysis presented in Section \ref{Rv_split_results}, included here for completeness. These figures respectively correspond to our models which allow for an intrinsic magnitude offset and a difference in baseline intrinsic SED between SNe Ia in high- and low-mass bins, with a split at 10$^{10}$ M$_\odot$. Please note that posterior samples of the difference between parameters in each mass bin are calculated simply by taking the difference between the two at each step along the MCMC chain.

\section{Post-processing Mass Split Model with Intrinsic SED Differences}
\label{appendix:postprocessing}

Equation \ref{bayesn_equation} gives an expression for the BayeSN model. It should be noted that if we apply the following transforms to the BayeSN model,

\begin{equation}
    \theta_1^s \rightarrow \theta_1^s - A \\
    W_0(t,\lambda_r) \rightarrow W_0(t,\lambda_r) + A W_1(t,\lambda_r)
\end{equation}
where A is an arbitrary constant, the resulting model is identical because the extra terms of $A W_1(t,\lambda_r)$ will cancel. As a result, a shift in $W_0(t,\lambda_r)$ which is proportional to $W_1(t,\lambda_r)$ will correspond to a shift in the mean of the $\theta_1$ distribution, which can be set arbitrarily. When the model is trained, we place a prior $\theta_1^s \sim N(0, 1)$ and the model will create its own consistent definition of $\theta_1$ relative to $W_0(t,\lambda_r)$ across the SN population.

However, a complication arises when allowing for intrinsic SED differences between SNe Ia in different mass bins as in Section \ref{binned_model_overview}. In the BayeSN model, $\theta_1$ acts as a stretch parameter representing the evolutionary timescale of the SN around peak. It is well established that SNe in higher mass galaxies have higher stretch values; we expect the mean of the $\theta_1$ distribution for the high-mass bin, $\Bar{\theta}_{1,\text{HM}}$, to be greater than the mean for the low-mass bin, $\Bar{\theta}_{1,\text{LM}}$. However, when we allow $W_0(t,\lambda_r)$ to vary between mass bins we are also implicitly allowing for the mean $\theta_1$ value to vary between mass bins, because of the transformation properties described above. Since we place a prior $\theta_1^s \sim N(0,1)$, $\Delta W^m$ will shift to separately set the mean of the $\theta_1$ distribution for each bin to be close to zero. However, we know that there is a physical difference in the mean $\theta_1$ between the two bins so $\theta_1=0$ in the high-mass bin will not correspond to the same light curve shape as $\theta_1=0$ in the low-mass bin.

The easiest way to consider this difference is to relate it to an observable property e.g.~$\Delta m_{15}$. We want a given value of $\theta_1$ to correspond to a given value of $\Delta m_{15}$, regardless of whether the SN is in the high-mass bin or the low-mass bin. In the $W_0$ split model, this is no longer necessarily the case. To give an example of why this comparison won't work, consider the following example:

\begin{itemize}
    \item In a model without a $W_0$ split, consider one SN in each mass bin. The SN in the low-mass bin has a latent $\theta_1^s=-0.3$, corresponding to $\Delta m_{15}=0.7$ mag. The SN in the high-mass bin, meanwhile, has a latent $\theta_1^s=0.0$, corresponding to $\Delta m_{15}=0.8$ mag. Of course, in reality, we have posterior distributions rather than fixed values, but for the sake of simplicity in this example we consider fixed values.
    \item Now we switch to a model which does have a $W_0$ split. As we have allowed $W_0$ to vary, there have been corresponding linear shifts in the $\theta_1$ distribution in each mass bin.
    \item For this new model, for the SN in the high-mass bin there has been no shift and we still infer $\theta_1^s=0.0$. However, for the low-mass bin there has been a shift and for the SN in this bin we now infer $\theta_1^s=0.0$, since the mean of the $\theta_1$ distribution has been increased by 0.3. Of course, this SN still has the same physical properties, with $\Delta m_{15}=0.7$ mag.
    \item If we are to naively evaluate the baseline intrinsic SED in each mass bin by setting $\theta_1^s=A_V^s=\epsilon^s(t,\lambda_r)=0$, we are in fact comparing the SED for two different stretch values -- we are comparing a SN with $\Delta m_{15}=0.7$ mag in the low-mass bin with a SN from the high-mass bin that has $\Delta m_{15}=0.8$ mag, not making a like-for-like comparison.
\end{itemize}

To ensure a fair comparison between high- and low-mass bins, we must apply some post-processing to the MCMC chains for $\Delta W^m$ to ensure that a single value of $\theta_1^s$ maps to the same value of $\Delta m_{15}$ for both mass bins. As previously shown, the model is invariant under a linear shift in $\theta_1$ and corresponding shift in $W_0$, therefore we can adjust $W_0$ to ensure this condition is met. For this work, we adjust $\Delta W^m$ such that for both mass bins $\Delta m_{15}$ in $g$-band $\Delta m_{g,15}=0.8$ mag. This choice is arbitrary and does not impact our results.

\begin{itemize}
    \item For each mass bin, we compute an updated $W^m_0(t,\lambda_r)=W_0(t,\lambda_r)+\Delta W^m(t,\lambda_r)$ where $W_0(t,\lambda_r)$ is the original $W_0$ matrix for the T21 BayeSN model.
    \item We use $W^m_0(t,\lambda_r)$ to compute $\Delta m^m_{g,15}$ in $g$-band, then calculate the difference between this and our reference value of 0.8 mag.
    \item We calculate the gradient $\frac{\partial\Delta m_{g,15}}{\partial\theta_1}$, which depends only on $W_1$ and is therefore the same between both mass bins since this parameter is kept fixed.
    \item We use the difference and gradient from above to calculate the correction in $\theta_1$ which will ensure $\Delta m_{g,15}(\theta_1=0)=0.8$ mag. Overall, for each mass bin we shift the $\theta_1$ distribution such that $\theta_1^{s,m} \rightarrow \theta_1^{s,m} - \theta^m_{1,\text{corr}}$ and in turn $W_0^m(t,\lambda_r) \rightarrow W^m_0(t,\lambda_r) + \theta^m_{1,\text{corr}}W_1(t,\lambda_r)$
\end{itemize}

The gradient $\frac{\partial\Delta m_{g,15}}{\partial\theta_1}$ can be calculated simply using the autodiff functionality of \textsc{jax}. Alternatively, an analytic expression can be derived as shown in Appendix \ref{appendix:deriv}. Strictly speaking, this gradient is weakly dependent on $\theta$, however in practice the variation is negligible over a reasonable range of $\theta$ values and we used the value for the model from \citet{T21} of $\frac{\partial\Delta m_{g,15}}{\partial\theta_1}|_{\theta_1=0}=0.1408$\footnote{After post-processing, at $\theta_1=0$ the values of $\Delta m^m_{g,15}$ are consistent for each mass bin to approximately 0.01 per cent.}. The expression for the correction factor on $\theta_1$ outlined above is therefore
\begin{equation}
    \theta^m_{1,\text{corr}} = \frac{\Delta m^m_{g,15}(\theta=0) - 0.8}{0.1408}
\end{equation}

This process is followed for every step along the chain to give a posterior distribution on $W^m_0(t,\lambda_r)$ with a consistent definition of $\theta_1$ across both bins. We can then set $\theta_1^s=A_V^s=\epsilon^s(t,\lambda_r)=0$ to do a like-for-like comparison between the baseline intrinsic SED for each mass bin.

\section{Relating the 15 day Decline in mag to BayeSN's Shape Parameter}
\label{appendix:deriv}

In this appendix we will derive an expression for the derivative of $\Delta m_{g,15}=m_g(t=15~\text{d})-m_g(t=0~\text{d})$, with respect to \textsc{BayeSN}'s light curve shape parameter $\theta_1$. Although we will write down this expression for the $g$-band, it can be applied to any passband.

The $g$-band apparent magnitude of a supernova at a rest frame phase $t$ is given by
\begin{equation}
    m_g(t) = -2.5\log_{10}[f_g(t)] + Z
\end{equation}
where $Z$ is the zero-point, and $f_g(t)$ is the \textsc{BayeSN} model flux (see \citealp{M20} eq.\ 4 and 6). The decline in rest-frame $g$-band magnitude in the 15 days after peak is given by
\begin{align}
    \Delta m_{g,15}&=m_g(t=15~\text{d})-m_g(t=0~\text{d})\\ &= -2.5\log_{10}\left[\frac{f_g(t=15~\text{d})}{f_g(t=0~\text{d})}\right]\\
    &=-\frac{2.5}{\ln 10}\ln\left[\int_{\lambda\in g}S(t=15~\text{d},\lambda)\mathbb{B}_g(\lambda)\,\lambda\,\text{d}\lambda\right]\nonumber\\ 
    &\quad+ \frac{2.5}{\ln 10}\ln\left[\int_{\lambda\in g}S(t=0~\text{d},\lambda)\mathbb{B}_g(\lambda)\,\lambda\,\text{d}\lambda\right],
\end{align}
where $\lambda$ is rest-frame wavelength, $\mathbb{B}_g(\lambda)$ is the transmission function for the $g$-band (normalised as in eq.\ 3 of \citealp{M20}), and $S(t,\lambda)$ is the \textsc{BayeSN} model SED. With no dust extinction ($A_V=0$) and no residual variation ($\epsilon(t,\lambda)=0$), the model SED is given by
\begin{equation}
    S(t,\lambda) = S_0(t,\lambda)10^{-0.4[M_0 + W_0(t,\lambda) + \theta_1W_1(t,\lambda)]}.
\end{equation}
A factor of $10^{-0.4M_0}$ can easily be cancelled from both integrals in the expression for $\Delta m_{g,15}$. We are interested in the gradient of the $\Delta m_{g,15}$ vs.\ $\theta_1$ relation, i.e.\ $d(\Delta m_{g,15})/d\theta_1$. Differentiating one of the terms in the expression for $\Delta m_{g,15}$ w.r.t.\ $\theta_1$, we find
\begin{equation}
    \frac{\text{d}}{\text{d}\theta_1}\ln\left[\int_{\lambda\in g}S(t,\lambda)\mathbb{B}_g(\lambda)\,\lambda\,d\lambda\right]=\frac{\int_{\lambda\in g}\frac{\partial S(t,\lambda)}{\partial\theta_1}\mathbb{B}_g(\lambda)\,\lambda \,\text{d}\lambda}{\int_{\lambda\in g}S(t,\lambda)\mathbb{B}_g(\lambda)\,\lambda\,\text{d}\lambda},
\end{equation}
where
\begin{equation}
    \frac{\partial S(t,\lambda)}{\partial\theta_1} = -\frac{\ln10}{2.5}W_1(t,\lambda)S(t,\lambda).
\end{equation}
Therefore, the derivative of interest is given by
\begin{align}
    \frac{\text{d}(\Delta m_{g,15})}{\text{d}\theta_1} &= \frac{\int_{\lambda\in g}W_1(t=15~\text{d},\lambda)S(t=15~\text{d},\lambda)\mathbb{B}_g(\lambda)\,\lambda\,\text{d}\lambda}{\int_{\lambda\in g}S(t=15~\text{d},\lambda)\mathbb{B}_g(\lambda)\,\lambda\,\text{d}\lambda}\nonumber\\
    &\quad-\frac{\int_{\lambda\in g}W_1(t=0~\text{d},\lambda)S(t=0~\text{d},\lambda)\mathbb{B}_g(\lambda)\,\lambda\,\text{d}\lambda}{\int_{\lambda\in g}S(t=0~\text{d},\lambda)\mathbb{B}_g(\lambda)\lambda\,\text{d}\lambda}
\end{align}


\bsp	
\label{lastpage}
\end{document}